\documentclass[twocolumn,amsmath,amssymb,floatfix,superscriptaddress,prl]{revtex4-2}
\usepackage{bm,graphicx,url,epsf,color}
\usepackage{amssymb}
\usepackage{amsthm}
\usepackage[colorlinks=true,linkcolor=blue,citecolor=blue]{hyperref}
\newcommand{\be}{\begin{equation}}
\newcommand{\ee}{\end{equation}}
\newcommand{\beS}{\begin{equation*}}
\newcommand{\eeS}{\end{equation*}}
\newcommand{\bea}{\begin{eqnarray}}
\newcommand{\eea}{\end{eqnarray}}
\newcommand{\ba}{\begin{eqnarray*}}
\newcommand{\ea}{\end{eqnarray*}}
\newenvironment{eqs}
{\begin{equation} \begin{aligned}}
{\end{aligned} \end{equation} }
\newcommand{\bal}{\begin{eqs}}
\newcommand{\eal}{\end{eqs}}

\newcommand{\bas}{\begin{eqs}}
\newcommand{\eas}{\end{eqs}}

\newcommand{\bK}{\mathbf{K}}
\newcommand{\bq}{\mathbf{q}}

\newcommand{\bk}{\mathbf{k}}

\newcommand{\bp}{\mathbf{p}}

\newcommand{\br}{\mathbf{r}}

\newcommand{\ket}[1]{\mid\! #1\rangle}
\newcommand{\bra}[1]{\langle #1\!\mid} 
\newcommand{\bw}{\begin{widetext}}
\newcommand{\ew}{\end{widetext}}

\newcommand{\ep}{{\epsilon}}

\begin{document}

\title{Higher-order van Hove singularity in magic-angle twisted trilayer graphene}
\author{Daniele Guerci}
\affiliation{Universit{\'e} de Paris, Laboratoire Mat{\'e}riaux et Ph{\'e}nom{\`e}nes Quantiques, CNRS, F-75013 Paris, France.} 
\affiliation{Universit{\'e} Paris-Saclay, CNRS, Laboratoire de Physique des Solides, 91405, Orsay, France.} 
\affiliation{Center for Computational Quantum Physics, Flatiron Institute, New York, New York 10010, USA}
\author{Pascal Simon}
\affiliation{Universit{\'e} Paris-Saclay, CNRS, Laboratoire de Physique des Solides, 91405, Orsay, France.} 
\author{Christophe Mora}
\affiliation{Universit{\'e} de Paris, Laboratoire Mat{\'e}riaux et Ph{\'e}nom{\`e}nes Quantiques, CNRS, F-75013 Paris, France.} 
\affiliation{Dahlem Center for Complex Quantum Systems and Fachbereich Physik, Freie Universit\"at Berlin, 14195, Berlin, Germany.}

\begin{abstract}

Twisted trilayer graphene (TTG) has recently emerged experimentally as a fascinating playground for studying correlated and exotic superconducting phases. 
We have found that TTG hosts a zero-energy higher-order van Hove singularity with an exponent $-1/3$ that is stronger than the one predicted in twisted bilayer graphene. 
This singularity is protected by a threefold rotation symmetry and a combined mirror-particle-hole symmetry and can be tuned with only the twist angle and a perpendicular electric field.
It arises from the combined merging of van Hove singularities and Dirac cones at zero energy, a scheme that goes beyond the recent classifications of van Hove singularities in single-band models. This new structure gives a topological Lifshitz transition, with anomalous exponent $-2/5$, which can be achieved in TTG by varying a third control parameter such as the atomic corrugation.  
The interplay between the non-standard class of higher-order van Hove singularities and interaction effects offers an unprecedented platform for studying correlation and superconductivity.

\end{abstract}

\date{\today}

\maketitle

\textit{Introduction.---}  Strong correlations generally result from a quenching of electronic motion, comparatively magnifying the strength of electron-electron interactions. This is the case for engineered flat bands, such as for fractional quantum Hall states, but also more generally in proximity to a singularity in the density of states. Tuning the chemical potential at a van Hove singularity (VHS) \cite{VanHove_1953,Lifshitz1960} introduces a large number of single-particle states with negligible energy likely to form a correlated state. Conventional van Hove singularities entail a logarithmic singularity but there are also higher-order types~\cite{shtyk2017,Liang_Fu_2019} with more diverging power law scaling which have been recently classified in single-band electron models~\cite{Liang_Fu_PRB_2020,Chamon_2020}. Such strong divergence amplifies correlation effects and plays a key role in determining the ordering instabilities in various materials, such as twisted bilayer graphene~\cite{Liang_Fu_2019,Betouras_2018,Isobe_2018,Yu_Ping_2020,Classen_2020,Chubukov_2020_TBG}, biased Bernal stacked bilayer graphene~\cite{shtyk2017}, twisted bilayer transition metal dichalcogenides~\cite{bi2020excitonic}, Sr3Ru2O7~\cite{Efremov_2019}, heavy fermions materials~\cite{Tsvelik_2012}, and high-${\rm T}_c$ superconductors~\cite{PhysRevLett.73.3302}. Moreover, it has been recently shown that the higher-order van Hove singularity gives rise to a novel non-Fermi liquid critical state dubbed supermetal~\cite{Isobe_2019}.

Moiré potentials obtained in graphene multi-layer structures by slight misalignment of the stacked layers have proven remarkably fruitful for tuning the single-particle spectrum~\cite{Castro_Neto_2007,Bistritzer_2011,Trambly_de_Laissardi_re_2010} and thereby achieving exotic phases driven by the combined effects of electronic correlation and topology~\cite{Cao_2018_ins,Cao_2018}. Twisted bilayer graphene (TBG) with two rotated graphene sheets thus exhibits a plethora of interesting phases~\cite{Andrei_2020}, including correlated symmetry breaking insulators~\cite{Jiang_2019,Zondiner_2020,Saito_2021,park2020flavour,Wong_2020,Xie_2019}, signatures of fragile topology~\cite{Song_2019,Vishwanath_2019_fragile_top}, orbital magnetism~\cite{XiDai_PRX_2019,Li_PRB_2020,Lu_2019,tschirhart2020imaging,guerci2021moire} and Chern insulators~\cite{Wu_2021,choi2020tracing,Nuckolls_2020,stepanov2020competing} with a quantum anomalous Hall effect~\cite{Sharpe_2019,Serlin_2020,Polshyn_2020}, nematicity~\cite{Choi_2019,Cao_2021}, and superconductivity~\cite{Cao_2018,Yankowitz_2019,Lu_2019}. Significant theoretical progress has been also achieved, especially in understanding the competing non-superconducting phases, see for instance Refs.~\cite{Zhang_2019,PhysRevX.9.021013,PhysRevLett.122.246401,PhysRevLett.122.246402,PhysRevLett.124.187601,PhysRevLett.124.166601,PhysRevB.102.235101,PhysRevB.102.035136,PhysRevB.103.035427,PhysRevLett.124.097601,PhysRevX.10.031034,PhysRevResearch.3.013033,Bernevig_2021_I,Song_2021_II,Bernevig_2021_III,Lian_2021_IV,Bernevig_2021_V,Xie_2021_VI}. A new appealing direction has been recently opened with experiments on twisted trilayer graphene (TTG)~\cite{Khalaf_2019,Christophe_2019,li2019electronic,PhysRevLett.125.116404,tsai2020correlated,Carr_2020,calugaru2021tstg,shin2021stacking,ramires2021emulating,Lopez_Bezanilla_2020,lei2021mirror,christos2021correlated,lake2021reentrant,qin2021inplane,fischer2021unconventional,chou2021correlationinduced}  where only the intercalated layer is rotated by a small angle. Convincing signatures of correlated phases and superconductivity have been observed~\cite{Park_2021,Hao_2021}, tunable with a perpendicular electrical (displacement) field. Interestingly, data suggest an unconventional superconducting state~\cite{cao2021large} in the strong coupling regime of tightly bound pairs and triplet, possibly p-wave, pairing.

In this letter, we argue that a strong higher-order van Hove singularity emerges in the single-particle spectrum of TTG upon tuning the displacement field and rotation angle. 
It results from the symmetric merging at zero energy of two standard VHS with opposite energies. 
Located at the band touching $\bK$ point of the Moiré Brillouin zone, it falls outside the single-band classification of VHS performed in Refs.~\cite{Liang_Fu_PRB_2020,Chamon_2020}, and also differs from the higher-order VHS identified~\cite{Liang_Fu_2019} in TBG. It exhibits the power law scaling $\omega^{-1/3}$, stronger than  $\omega^{-1/4}$  predicted for TBG. 
An even stronger exponent $-2/5$ is found by tuning the corrugation, indicating a Lifshitz transition between two topologically incompatible energy contours.


\textit{Model and mapping to TBG.---} The starting point is the continuum model~\cite{Castro_Neto_2007,Bistritzer_2011} for trilayer graphene, where the three layers are stacked with alternating twist angles $\pm\theta$~\cite{Khalaf_2019}, coupling the three Dirac cones in each valley.  It is convenient to take advantage of the mirror symmetry $M_z$ with respect to the middle layer in the absence of displacement field $D/\varepsilon_0$, and write the Hamiltonian in an already layer-rotated basis~\cite{Khalaf_2019,Carr_2020,PhysRevLett.125.116404,calugaru2021tstg}
\begin{equation}
\label{H_STTG}
H(\br)=\left(\begin{array}{ccc} \hbar v_0\bm{\sigma}\cdot\hat{\bk} & \sqrt{2}\,T(\br) & 0 \\ \sqrt{2}\,T^\dagger(\br) & \hbar v_0\bm{\sigma}\cdot\hat{\bk} & U\,\sigma^0/2 \\  0 & U\,\sigma^0/2 & \hbar v_0\bm{\sigma}\cdot\hat{\bk}\end{array}\right),
\end{equation}
where $\hat{\bk}=-i\nabla_{\br}$, $v_0$ is the electrons velocity in graphene, $T(\br)=\sum_{j=1}^{3}e^{i\bq_j\cdot\br}\,T_j$ are the interlayer hoppings with sublattice structure, $\bq_1=k_\theta(0,1)$, $\bq_{2/3}=k_\theta(\mp\sqrt{3}/2,-1/2)$, where $k_\theta = |{\bf K}| \theta$ is the distance between the ${\bf K}$ points of consecutive layers.
$\pm U$ are the gate potentials applied on the top and bottom layers, $U = d D/\varepsilon_0 \varepsilon$, where $d\simeq 0.3\,\text{nm}$ is the interlayer distance and $\ep_0$ is the bare dielectric constant. For simplicity we neglect screening of the displacement and take $\varepsilon=1$. The interlayer tunneling matrices take the form 
$T_{j+1}=w_{AA}\sigma^0+w_{AB} \left[ \sigma^+ e^{-2 i \pi j /3} + \sigma^- e^{2 i \pi j /3} \right]$,
where $j=0,1,2$, $w_{AB}=w$ and $w_{AA}=w\,r$ with $r<1$ due to lattice relaxation effects~\cite{Koshino_2017,Koshino_2018}, and the $\sigma^{0,\pm}$ matrices act in sublattice space. Hereafter we take $r=0.8$ unless stated otherwise. As Eq.~\eqref{H_STTG} governs the valley ${\bf K}$,  the valley ${\bf K'}$ is simply obtained by time-reversal symmetry, {\it i.e.} complex conjugation.
\begin{figure}
\begin{center}
\includegraphics[width=0.49\textwidth]{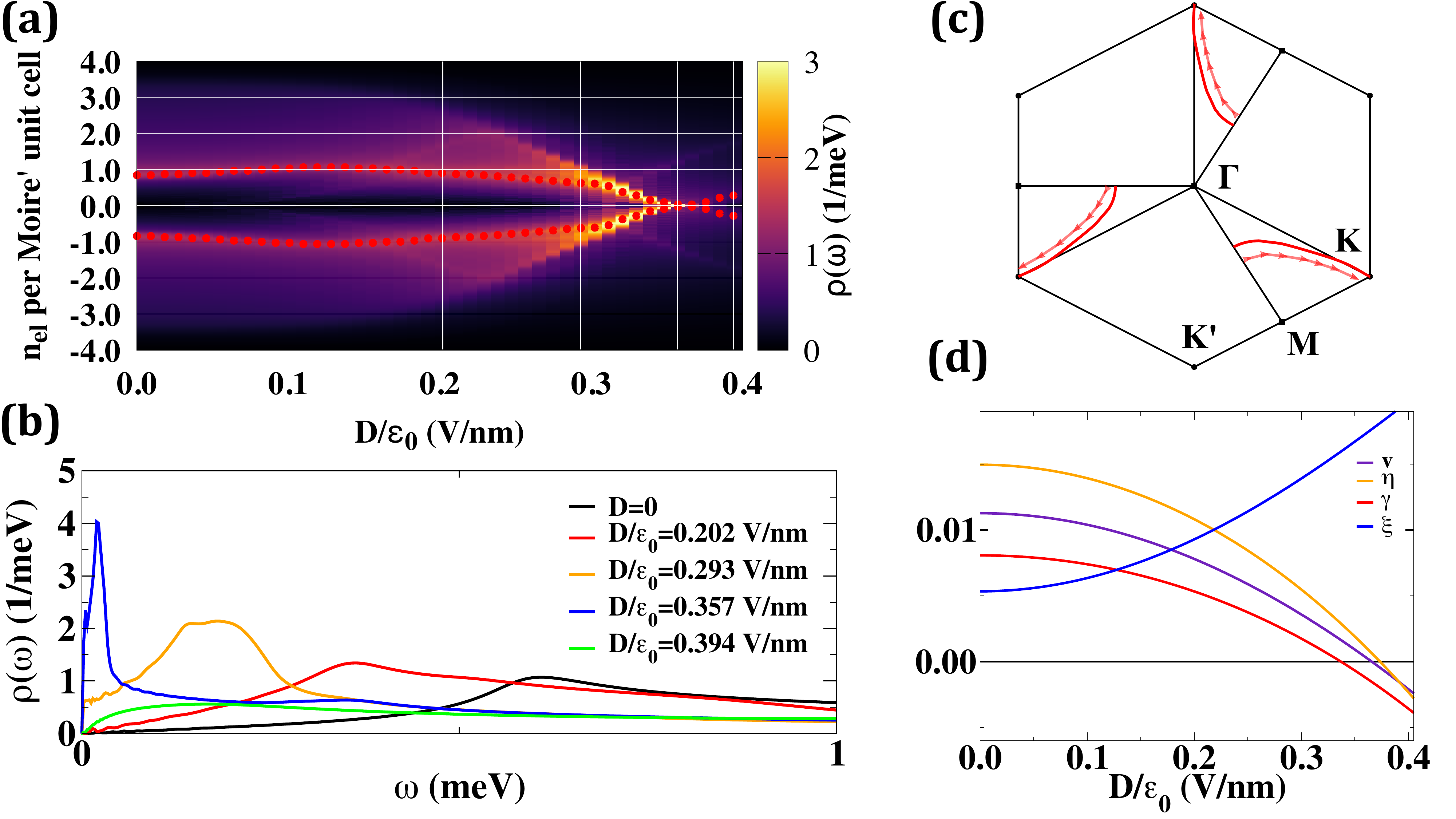}
\caption{(a) DOS as a function of the displacement field $D/\ep_0$ and of the number of electrons per Moir\'e unit cell for $r=0.8$ and $\theta=1.59^\circ$. The location of the two VHS are indicated by red dots. (b)
line cuts, corresponding to the white lines in (a), of the DOS as function of energy for various displacement fields. (c) Evolution as a function of the displacement field of the saddle points (red lines) yielding the strongest peak in the density of states. Arrows indicate increasing $D/\ep_0$. (d) Values of the parameters of the Hamiltonian~\eqref{k_dot_p_Hamiltonian} $H_\bK(\bq)$ as a function of $D/\ep_0$. }
\label{DOS_159_08}
\end{center}
\end{figure}

At zero displacement $U=0$, Eq.~\eqref{H_STTG} decouples a high-velocity Dirac cone in the odd mirror sector, located at the ${\bf K'}$ point of the Moiré Brillouin zone,  from two coupled Dirac cones in the even mirror sector, one at ${\bf K}$ and the other one at ${\bf K'}$. The even sector maps exactly onto the TBG Hamiltonian with a rescaled interlayer tunneling~\cite{Khalaf_2019}, and a corresponding $\sqrt{2}$ enhancement of the magic angle $\theta_m \simeq1.541^\circ$ at which the Dirac velocity vanishes. Computing the density of states, we recover the symmetric VHS of TBG in the two active bands, see Figs.\ref{DOS_159_08} (a) and (b), observed in STS experiments~\cite{Li_2009,PhysRevB.92.155409,Kim_2017,Kerelsky_2019,Wu_2021b}. The VHS is in fact composed by three equivalent saddle points related by $C_{3 z}$ symmetry, {\it i.e.} $2 \pi/3$ rotation in the graphene plane.
At a critical angle $\theta_v\simeq1.571^\circ$, a higher-order VHS~\cite{Liang_Fu_2019} occurs below which each saddle point splits in two. The DOS singularity at $\theta_v$ is a power law with exponent $-1/4$ and asymmetry ratio $\sim\sqrt{2}$, which convincingly matches the experimentally observed VHS peak~\cite{Kerelsky_2019}.
The $C_{2 x}$ symmetry denotes a $\pi$ rotation around $x$ exchanging layer and sublattice indices. It enforces the saddle points to be on the $\bm{\Gamma}\mathbf{M}$ lines of the Moiré Brillouin zone for $\theta$ larger than $\theta_v$. Throughout the rest of this paper, we will focus on angles $\theta$ such that the higher-order VHS inherited from TBG does not play a role. Aside from the symmetries $M_z$, $C_{3z}$, $C_{2 x}$ already introduced, the model in Eq.~\eqref{H_STTG} is also invariant under the combination $C_{2z}T$. It anticommutes with the particle-hole symmetry $P$ in the even mirror sector and with $C_{2 x} P$ in both odd and even subspaces.

\textit{VHS merging and effective model.---} We discuss the case of non-zero displacement field $U \ne 0$ and explore the evolution of the density of states at the twist-angle $\theta=1.59^\circ$. 
$U$ breaks the mirror symmetry $M_z$, the rotation symmetry $C_{2x}$ and the particle-hole symmetry $P$. However, the product $M_z C_{2x} P$ is preserved~\cite{calugaru2021tstg} together with the remaining symmetries. $M_z C_{2x} P$ acts as anticommuting particle-hole symmetry. It will be very important in stabilizing the new van Hove singularity, see below. Fig.~\ref{DOS_159_08} (a) shows the density of states as a function of the number of electrons per Moir\'e unit cell and electric displacement field $D/\varepsilon_0$. The two particle-hole symmetric VHS, represented as red dots in Fig.~\ref{DOS_159_08} (a),  move towards charge neutrality as $D$ increases and merge at zero energy at a critical $D_c/\varepsilon_0 \simeq 0.37\,\text{V/nm}$. The VHS peaks become concomitantly more pronounced with increasing $D$ as shown in Fig.~\ref{DOS_159_08} (b), moving towards what seems to be a zero-energy divergence. The corresponding evolution of the $C_{3 z}$ symmetric saddle points with the displacement field is shown in Fig.~\ref{DOS_159_08} (c). As a consequence of $C_{2 x}$ symmetry breaking, the saddle points leave the  $\bm{\Gamma}\mathbf{M}$ lines and converge towards the ${\bf K}$ points in the Moiré Brillouin zone. By further increasing $D$ above $D_c$ the VHS split again and move away from charge neutrality together with a substantial reduction of the VHS peaks.

In order to gain more analytical insight into the zero-energy merging of VHS and the marked singularity in the density of states, we derive a low-energy approach close to the ${\bf K}$ point in the Moiré Brillouin zone. Being a high-symmetry point, ${\bf K}$ retains some of the symmetries of the model that  leaves it invariant: $C_{3 z}$, $C_{2 z} T$ and $M_z C_{2x} P$. $C_{2 z} T$ protects a Dirac cone at ${\bf K}$ even at non-zero displacement field. The two degenerate states at ${\bf K}$, denoted $u_\omega$ and $u_{\omega^*}$, can be classified by their $C_{3 z}$ eigenvalues $\omega=e^{2i\pi/3}$ and $\omega^*$, respectively. $M_z C_{2x} P$ pins these two states at zero energy and more generally enforces a fully particle-hole symmetric spectrum at ${\bf K}$. The form of the low-energy Hamiltonian in the vicinity of ${\bf K}$ is constrained by the symmetries. In the basis of the two degenerate states $(u_\omega, u_{\omega^*})$ defining the Pauli matrices $\tau_{0,x,y,z}$, it takes the  $C_{3 z}$-symmetric form
\bal
\label{k_dot_p_Hamiltonian}
H_\bK (\bq)=&\hbar v  \,\bm{\tau}\cdot\bq+\eta(q^2_y-q^2_x)\tau_y-2\eta q_x q_y\tau_x\\
&+2\gamma(q^3_x-3q_xq^2_y)\tau_0+\xi\, q^2\,\bm{\tau}\cdot\bq,
\eal 
to third order in  $\bq = \bk - \bK$, where $q^2=q^2_x+q^2_y$ and the couplings $\{v,\eta,\gamma,\xi\}$ are calculated by employing a $\bk\cdot\bp$ approach with degenerate perturbation theory (see below). The symmetry $M_z C_{2x} P$, acting as $\tau_x H_\bK (q_x,q_y) \tau_x = -H_\bK (-q_x,q_y)$, forbids  other $C_{3 z}$-symmetric terms, such as $q^2 \tau_0$, $(q_x^2 -q_y^2) \tau_x - 2 q_x q_y \tau_y$  or $(q_y^3-3 q_y q_x^2) \tau_0$, to appear in Eq.~\eqref{k_dot_p_Hamiltonian}.
The crucial reduction to four coupling constants $\{v,\eta,\gamma,\xi\}$ implies that the higher-order VHS, taking place when $v$ and $\eta$ both vanish, requires the fine-tuning of only two parameters of the original model, such as the twist angle $\theta$ and the displacement field $D$.

\begin{figure}
\begin{center}
\includegraphics[width=0.485\textwidth]{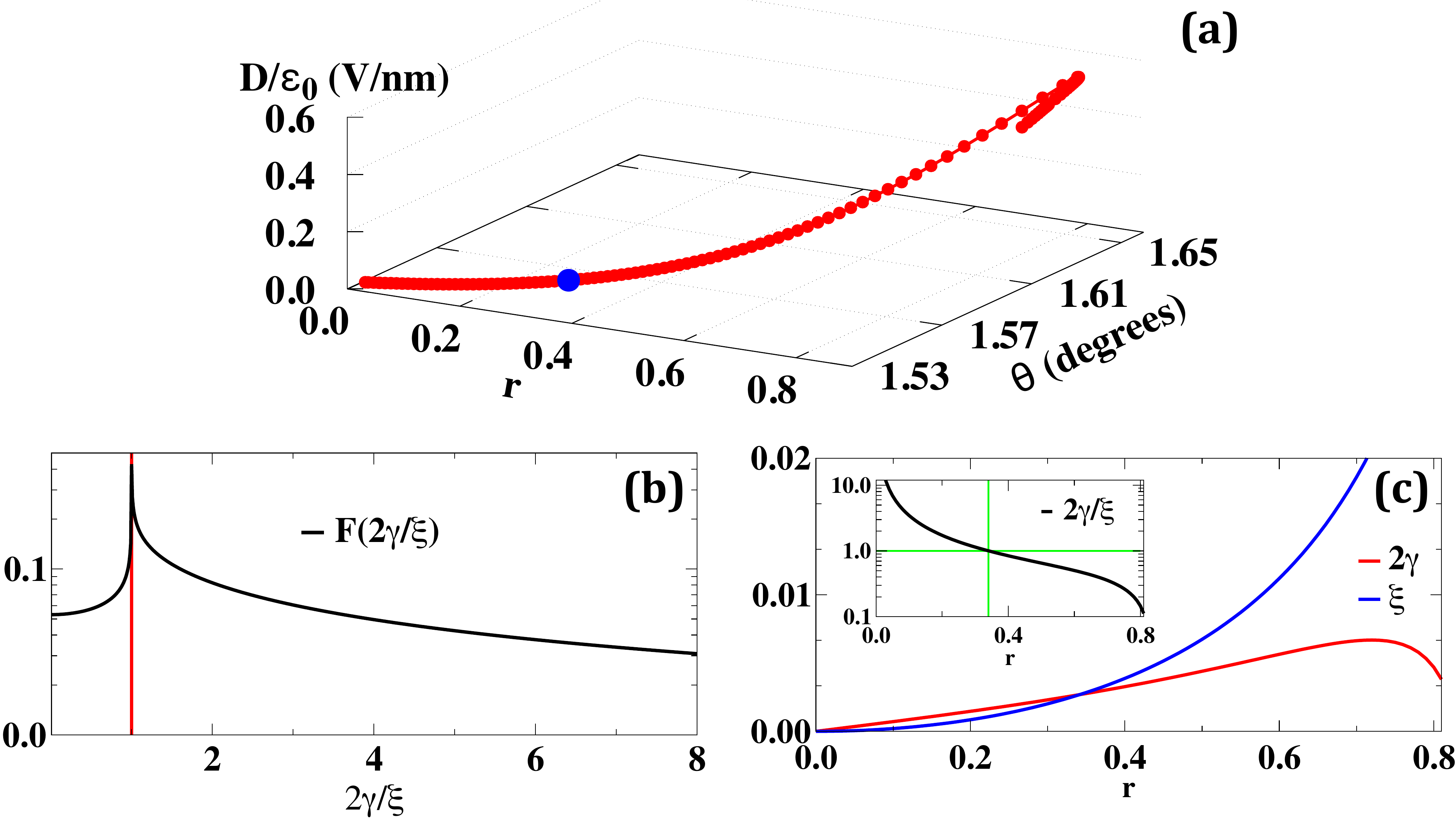}
\caption{(a) Three-dimensional trajectory of higher-order VHS originating from the first magic-angle $\theta_{CH}\simeq1.536^\circ$ at $r=0$ and $D=0$.  The blue dot gives the position of the Lifshitz transition with exponent $-2/5$. (b) Evolution of the DOS prefactor in Eq.~\eqref{DOS} as function of $2\gamma/\xi$. (c) Evolution of $2\gamma$ and $\xi$ as a function of $r$ along the trajectory of higher-order VHS. The inset shows the ratio $2\gamma/\xi$ in logarithmic scale.}
\label{trajectory_3D}
\end{center}
\end{figure}

Beyond the zero-energy subspace $(u_\omega, u_{\omega^*})$, it is convenient to introduce the other eigenstates of Eq.~\eqref{H_STTG} at $\bK$, $\ket{u_n}=\ket{u_{n\bK}}$  with energies $\epsilon_n$, and the operator $\mathcal{P}=-\sum_{n}\ket{u_n}\bra{u_n}/\epsilon_n$. The values of the coupling constants 
\begin{equation}
\label{parameters}
\begin{aligned}
&v=\bra{u_{\omega^*}}\sigma^-\ket{u_\omega},\\
&\eta=-\text{Im}\left[\bra{u_{\omega^*}}\sigma^+\,\mathcal{P}\,\sigma^+\ket{u_{\omega}}\right],\\
&\gamma=\text{Re}\left[\bra{u_{\omega}}\sigma^-\,\mathcal{P}\,\sigma^-\,\mathcal{P}\,\sigma^-\ket{u_{\omega}}\right],\\
&\xi=\text{Re}\Big[\bra{u_{\omega^*}}\sigma^+\,\mathcal{P}\,\sigma^-\,\mathcal{P}\,\sigma^-+\sigma^-\,\mathcal{P}\,\sigma^+\,\mathcal{P}\,\sigma^-\\
&+\sigma^-\,\mathcal{P}\,\sigma^-\,\mathcal{P}\,\sigma^+\ket{u_{\omega}}\Big],
\end{aligned}
\end{equation}
are calculated from a $\bk\cdot\bp$ approach at $\bK$~\footnote{See Supplemental Material for details of the calculations.} and depend on the details of the spectrum of the trilayer. They are shown in
 Fig.~\ref{DOS_159_08} (d) as a function of the displacement field for the angle $\theta=1.59^\circ$ measured in~\cite{Park_2021}. 
Remarkably, $v$ and $\eta$ vanish for values of $D$ in very close proximity, suggesting the vicinity to the higher-order VHS. It explains the strong feature observed Fig.~\ref{DOS_159_08}  in the density of states although, rigorously speaking, reaching the higher-order VHS requires the fine-tuning of an additional parameter. This can be done by changing the twist angle $\theta$.
In order to get a closer look at the exact position of the higher-order VHS, we trace out the values of $D$ and $\theta$ for which the velocity and the curvature of the band dispersion both vanish at the $\bK$ point. The result as  function of the corrugation parameter $r$ is shown in Fig.~\ref{trajectory_3D} (a).
As anticipated, we find a line of higher-order VHS in this tridimensional parameter space. We stress again that, under the reasonable assumption of a corrugation parameter $r=0.8$ and for the twist angle  $\theta=1.59^\circ$, one gets very close to the higher-order VHS by simply tuning the electric displacement. 

The trajectory of higher-order VHS in Fig.~\ref{trajectory_3D} (a) originates from a critical angle $\theta_{CH} \simeq 1.536^\circ$ where both the displacement field $D$ and the corrugation $r$ are vanishing. This point corresponds in fact to the chiral limit of TBG~\cite{Vishwanath_2018,PhysRevResearch.2.023237,Becker_2021,Wang_2021,Ren_2021} where the whole active band is rigorously flat.
At $r=0$, $\theta =\theta_{CH}$  is also the first magic angle. As $D$ increases, the line of higher-order VHS extends till a critical value of the atomic corrugation $r \simeq 0.842$ above which it is not possible to have $v$ and $\eta$ both vanishing.
At smaller twist angles, we find that each magic angle $\theta^{(n>1)}_{CH}$ in the chiral limit $r=0$ is the starting point of a similar line of higher-order VHS.

\textit{The higher-order van Hove singularity.---}  The higher-order VHS appears for a vanishing velocity $v=0$ and curvature $\eta=0$ in Eq.~\eqref{k_dot_p_Hamiltonian}. In this case,  the dimensional scaling $H_{\bK}(\lambda^{1/3} \bq)=\lambda \, H_{\bK}(\bq)$ indicates a power-law scaling
\begin{equation}\label{DOS}
    \rho(\omega) = \xi^{-2/3} \mathcal{F}(2\gamma/\xi) \,  |\omega|^{-1/3}
\end{equation}
close to charge neutrality, with exponent $-1/3$. The Hamiltonian describes a cubic band touching and therefore extends the one-band classification of VHS~\cite{Liang_Fu_PRB_2020,Chamon_2020}. We would recover effectively one band by setting $\xi=0$ and having $H_\bK (\bq)=
2\gamma(q^3_x-3q_xq^2_y)\tau_0 $. In the vicinity of the higher-order VHS, there are three $C_{3 z}$-symmetric saddle points at positive energy, three at negative, all merging at $\bK$ as $v$ and $\eta$ are tuned to zero, as shown in Fig.~\ref{DOS_159_08} (c), reminiscent of the saddle point merging in the single band case~\cite{shtyk2017}. In addition, the effective two-band structure entails a pseudospin with a vorticity of $+1$ at $\bK$ protected by $C_{2 z} T$. In the vicinity of the higher-order VHS, $\bK$ hosts a $+1$ Dirac cone surrounded by six additional Dirac cones~\cite{Note1}. The side cones are organized in two $C_{3z}$-symmetric triplets along $\bK^\prime\bK$ ($\bm{\Gamma}\bK$) with vorticity $+1$ ($-1$). The higher-order VHS occurs precisely as all Dirac cones meet at $\bK$, leading to cubic band touching with $+1$ vorticity. A single Dirac cone at $\bK$ remains above the critical value $D>D_c$.

\begin{figure}
\begin{center}
\includegraphics[width=0.49\textwidth]{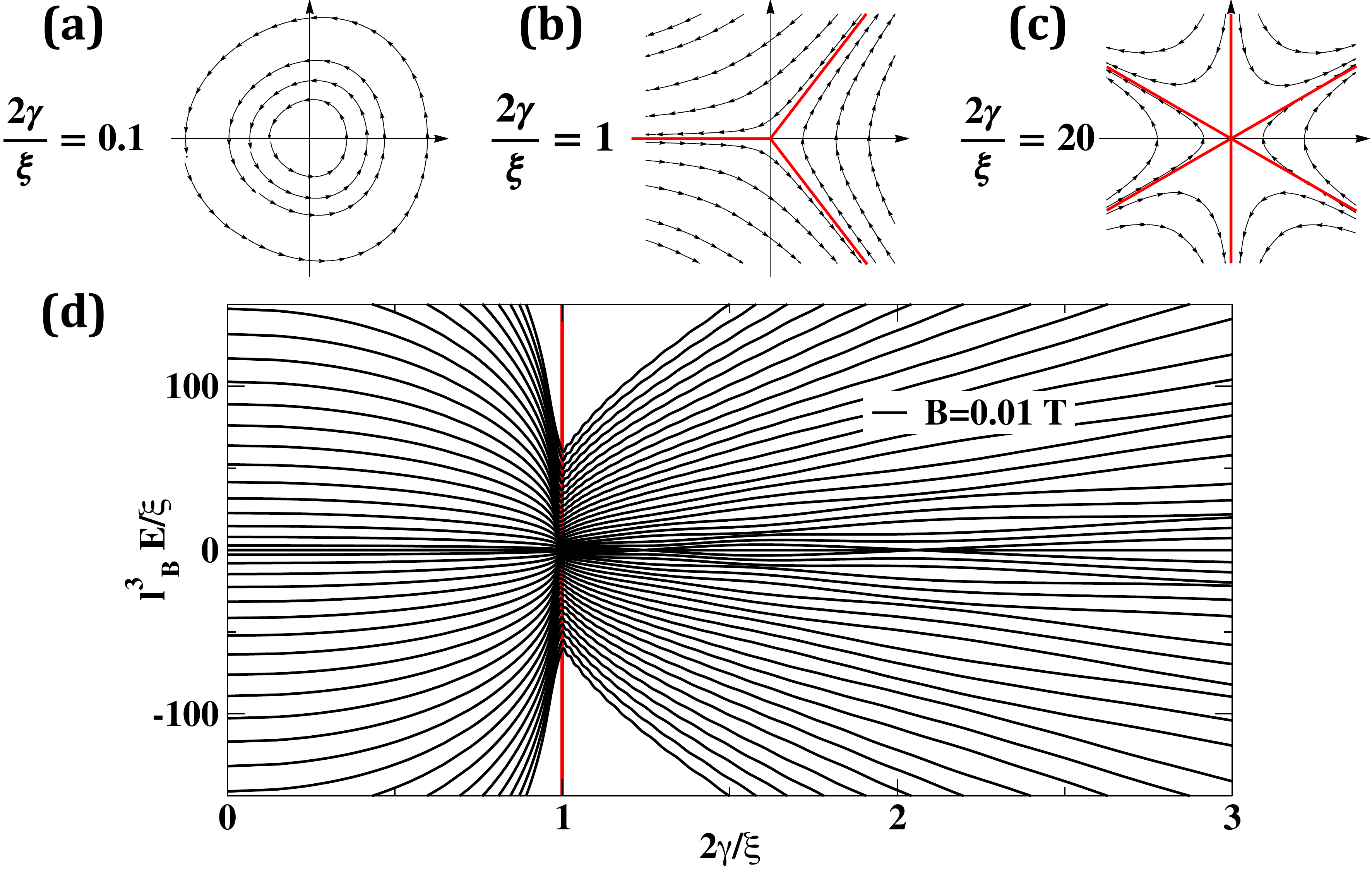}
\caption{(a-c) show the positive iso-energy contours around $\bK$ at the higher-order VHS for $2\gamma/\xi=0.1$, $2\gamma/\xi=1.0$ (Lifshitz transition) and $2\gamma/\xi=20.0$, respectively. Red lines mark the asymptotes of open orbits. (d) Landau level spectrum as a function of $2\gamma/\xi$ for $B=0.01\,\text{T}$. The vertical red line indicates the location of the Lifshitz transition. The Landau level energy is measured in unit of $\xi/l^3_B$ where $l_B$ is the magnetic length $l_B=\sqrt{\hbar/eB}$.   }
\label{LLs_Orbits}
\end{center}
\end{figure}

In the one-band limit of vanishing $\xi$, the semiclassical orbits are fully open with an anisotropic elliptic umbilic ($D_4^-$) structure~\cite{Liang_Fu_PRB_2020,Chamon_2020} separated by $C_{3 z}$-symmetric lines and a cusp at $\bq = 0$. Quite on the contrary, $\gamma=0$ predicts closed and isotropic semiclassical orbits. The corresponding iso-energy contours are illustrated in Fig.~\ref{LLs_Orbits} (a-c). This difference in topology indicates that a Lifshitz transition is expected to occur as function of the ratio $2\gamma/\xi$. This is visible in the divergence in the prefactor $\mathcal F$ of Eq.~\eqref{DOS} shown in Fig.\ref{trajectory_3D} (b).
Indeed, for $\xi = 2 \gamma$, the spectrum at the higher-order VHS is $E_{\pm}(\bq) =\xi q^3 [\pm 1 + \cos(3 \phi)]$, where $\phi$ is the polar angle between $\bq$ and the $x$-axis. It predicts zero-energy lines along the axis at $\phi =\pm \pi/3,\pi [\pi]$, shown in Fig.~\ref{LLs_Orbits} (b), which are responsible for an even stronger divergence in the density of states. The zero-energy lines are lifted by further expanding the $\bk\cdot\bp$ approach in Eq.~\eqref{k_dot_p_Hamiltonian} to the first non-vanishing order. The fourth order being zero, due to the symmetries $C_{3 z}$ and $M_z C_{2x} P$, the dispersion finally takes the form~\cite{Note1}
\begin{equation}
E_{+}(\bq) = \xi q^3 [1 + \cos(3 \phi)] + \Lambda q^5
\end{equation}
in the vicinity of the critical line at $\phi$ (the other lines are inferred by symmetry), where $\Lambda$ is a positive coefficient. 
Thanks to this term, the divergence in the density of states is cut off at the angle $\delta\phi \sim \sqrt{\Lambda/\xi}(\omega/\Lambda)^{1/5}$, where $\delta \phi = \phi - \pi$, and a new power law is obtained
\begin{equation}
\rho (\omega) \sim |\omega|^{-2/5},
\end{equation}
stronger than the naive scaling dimension $-1/3$ of the higher-order VHS. 

Coming back to the original trilayer model in Eq.~\eqref{H_STTG}, the ratio $2\gamma/\xi$ along the (red) line of higher-order VHS is represented in Fig.~\ref{trajectory_3D} (c). We then find that the critical point $\xi = 2 \gamma$ is reached when the corrugation parameter is $r \simeq 0.34$, corresponding to the twist angle $\theta \simeq 1.543^\circ$ and the displacement field $D/\varepsilon_0 \simeq 0.086\,\text{V/nm}$. At this point, the DOS critical exponent thus changes to $-2/5$ while it is $-1/3$ along the rest of the line, indicating a change of topology of the iso-energy contours, from local open orbits when $r<0.34$, to closed ones when $r>0.34$.

The change of Fermi surface topology is reflected in the Landau level (LL) spectrum \cite{Alexandradinata_2018} calculated in Fig.~\ref{LLs_Orbits} (d) for $0.01 \,\text{T}$. It is obtained from minimal coupling applied to the effective model in Eq.~\eqref{k_dot_p_Hamiltonian}, at vanishing $(v,\eta)$, where the $q^4$ terms have been added to regularize the region with open orbits. We find markedly different behavior of both sides of the Lifshitz transition: a regularly spaced structure for closed orbits, corresponding to $2\gamma/\xi<1$ (and $r>0.34$), which evolves into weak oscillations in the regime of open orbits. The LL energy scaling $\propto (n B)^{3/2}$~\cite{shtyk2017} of the higher-order VHS, with LL index $n$ and magnetic field $B$, becomes $\propto (n B)^{5/3}$ at the transition due the anomalous $-2/5$ DOS exponent. This yields in Fig.~\ref{LLs_Orbits} (d) a relative collapse of LL at the transition, even more pronounced for decreasing $B$.

\textit{Conclusions.---} 
We derived the existence of a higher-order VHS in mirror-symmetric TTG associated with a strong zero-energy peak in the density of states.
In contrast with twisted bilayer graphene, the HOVHS arises from the fusion between the standard finite energy VHS and the Dirac cone at $\bK$, and it is protected by $C_{3z}$ and $M_zC_{2x}P$.  
As long as these two symmetries are not broken, the HOVHS will still be present with effective parameters which are renormalized by the electron-electron interaction. We emphasize that the peak in the DOS is strongly enhanced in the vicinity of the HOVHS, as shown in Fig.~\ref{DOS_159_08}(a), and that the position shifts toward charge neutrality. This has two consequences: there will be a larger range of doping for which the Stoner criterion is fulfilled, and this criterion is satisfied for smaller doping to charge neutrality.
We argued that current experiments at the magic angle can be brought close to this strong singularity by electric gating, although very strong fields might be needed to compensate screening. 
This finding has far-reaching consequences, as this strong divergence in the density of states, will result in the emergence of a plethora of many-body phenomena that have barely begun to be studied~\cite{christos2021correlated}.
Moreover, the type of van Hove singularity that we found differs fundamentally from the standard van Hove singularity, such as the one classified in Refs.~\cite{Liang_Fu_PRB_2020,Chamon_2020}, as it also involves a Dirac point. This new structure gives a topological Lifshitz transition with anomalous exponent $-2/5$ that in the regime of small magnetic fields results in a Landau level energy collapse. Our work opens a new horizon in the study of the interplay between new correlated or superconducting phases and exotic van Hove singularities that are realized in Moir\'e superlattice materials.

\textit{Acknowledgments.---} We have benefited from discussions with K. Kol\'{a}\v{r} and F. von Oppen. This work was supported by the French National Research Agency (project  SIMCIRCUIT, ANR-18-CE47-0014-01). The Flatiron Institute is a division of the Simons Foundation.


\bibliography{mybiblio}
\end{document}


\title{Supplementary Material: Higher-order van Hove singularity in magic-angle twisted trilayer graphene}
\author{Daniele Guerci}
\affiliation{Universit{\'e} de Paris, Laboratoire Mat{\'e}riaux et Ph{\'e}nom{\`e}nes Quantiques, CNRS, F-75013 Paris, France} 
\affiliation{Universit{\'e} Paris-Saclay, CNRS, Laboratoire de Physique des Solides, 91405, Orsay, France.} 
\affiliation{Center for Computational Quantum Physics, Flatiron Institute, New York, New York 10010, USA}
\author{Pascal Simon}
\affiliation{Universit{\'e} Paris-Saclay, CNRS, Laboratoire de Physique des Solides, 91405, Orsay, France.} 
\author{Christophe Mora}
\affiliation{Universit{\'e} de Paris, Laboratoire Mat{\'e}riaux et Ph{\'e}nom{\`e}nes Quantiques, CNRS, F-75013 Paris, France.} 
\affiliation{Dahlem Center for Complex Quantum Systems and Fachbereich Physik, Freie Universit\"at Berlin, 14195, Berlin, Germany.}

\date{\today} 
\pacs{}

\maketitle

\section{The continuum model for STTG}
\label{sec1}

\begin{figure}
\begin{center}
\includegraphics[width=0.6\textwidth]{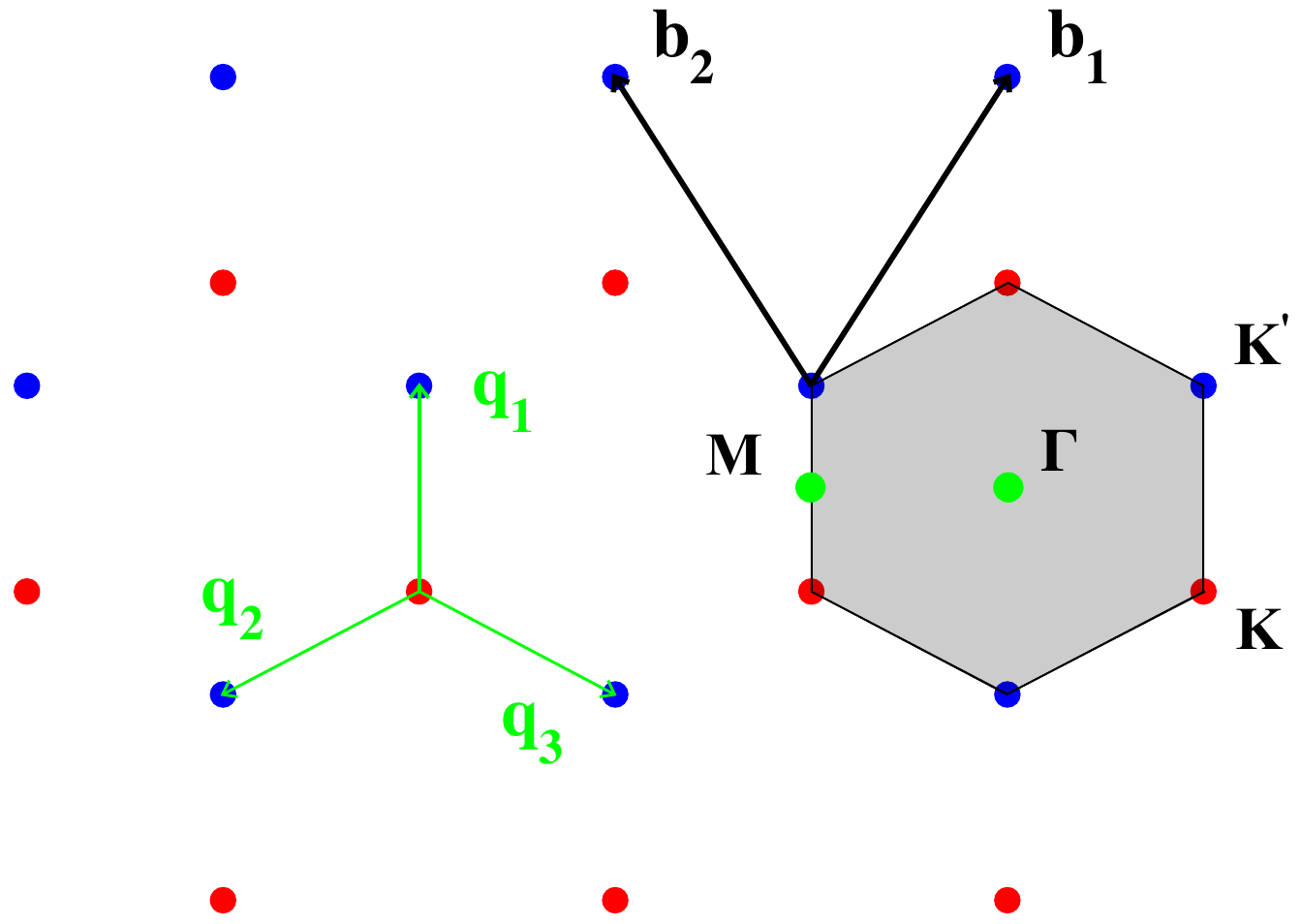}
\caption{the $\bk$-space lattice generated by $\bq_1$, $\bq_2$ and $\bq_3$. The honeycomb lattice is composed by two interpenetrating triangular sublattices $\mathcal{Q}_+$ and $\mathcal{Q}_-$ corresponding to the middle (red dots) and the bottom/top (blue dots) layers, respectively. The grey shaded region centered at $\bm{\Gamma}$ depicts the MBZ, $\bK$ and $\bK^\prime$ are the high-symmetry points at the corner of the MBZ, while the $\mathbf{M}$ point is the middle point between $\bK$ and $\bK^\prime$. The vectors $\mathbf{b}_1=\bq_1-\bq_2$ and $\mathbf{b}_2=\bq_1-\bq_3$ are the primitive vectors of the triangular lattice. Given $\mathcal{Q}_0=\mathbb{Z}\mathbf{b}_1+\mathbb{Z}\mathbf{b}_2$ with $\mathbb{Z}$ the set of relative integers, the two sublattices $\mathcal{Q}_+$ and $\mathcal{Q}_-$ are obtained as $\mathcal{Q}_+=\mathcal{Q}_0+\bq_2$ and $\mathcal{Q}_+=\mathcal{Q}_0-\bq_3$.}
\label{R_lattice_TTG}
\end{center}
\end{figure}

We use an effective continuum model for the low-energy electronic bands of the symmetric twisted trilayer graphene. 
Following Refs. \cite{Khalaf_2019,Carr_2020,PhysRevLett.125.116404,calugaru2021tstg} the Hamiltonian reads:
\be
\label{H_STTG}
{H}(\br)=\left(\begin{array}{ccc} \hbar v_0\,\bm{\sigma}_{\theta/2}\cdot\hat{\bk}+U\,\sigma^0/2 & T^\dagger(\br) & 0 \\ T(\br) & \hbar v_0\bm{\sigma}_{-\theta/2}\cdot\hat{\bk} & T(\br) \\ 0 & T^\dagger(\br) & \hbar v_0\bm{\sigma}_{\theta/2}\cdot\hat{\bk}-U\,\sigma^0/2\end{array}\right),
\ee
where $\hat{\bk}=-i\nabla_\br$, $T(\br)=\sum_{j=1}^{3}\,T_j\,e^{i\bq_j\cdot\br}$, $\bq_1=k_\theta(0,1)$, $\bq_{2/3}=k_\theta(\mp\sqrt{3}/2,-1/2)$, $k_\theta=4\pi/3\,L_s$ with $a$ the graphene lattice parameter and $L_s=a/(2\sin\theta/2)\simeq a/\theta$ the linear size of the Moir\'e unit cell, $U$ is the gate potential applied on the top and bottom layers and $\sigma^i_{\pm\theta/2}=\sum_{j=x,y}R_{ij}(\pm\theta/2)\,\sigma^j$ with $R_{ij}(\phi)$ rotation of angle $\phi$ around the $\bm{z}$ axis.
In the Hamiltonian \eqn{H_STTG} we have also introduced the interlayer tunneling matrix 
\be
T_{j+1}=w_{AA}\sigma^0+w_{AB}\left[\cos\frac{2\pi\,j}{3}\sigma^x+\sin\frac{2\pi\,j}{3}\sigma^y\right],
\ee 
where $j=0,1,2$, $w_{AB}=w$ and $w_{AA}=w\,r$ with $r<1$ due to lattice relaxation effects \cite{Koshino_2017,Koshino_2018}.
In the absence of the displacement field $U=0$ the Hamiltonian \eqn{H_STTG} is invariant under the unitary transformation: 
\be
M_z=\left(\begin{array}{ccc}0 & 0 & \sigma^0 \\0 &  \sigma^0 & 0 \\ \sigma^0 & 0 & 0\end{array}\right),
\ee
which leaves the middle layer invariant while exchanging bottom and top layers. The mirror symmetry $M_z$ is characterized by two eigenstates $\ket{e_2}=(\ket{1}+\ket{3})/\sqrt{2}$ and $\ket{e_1}=\ket{2}$ with eigenvalue $+1$, and one eigenstate $\ket{o}=(\ket{1}-\ket{3})/\sqrt{2}$ with $-1$. 
In the basis of the eigenstates of $M_z$ the Hamiltonian $H(\br)$ becomes: 
\be
\label{H_STTG_NB}
W^\dagger\,{H}(\br)\,W=\left(\begin{array}{ccc} \hbar v_0\bm{\sigma}_{-\theta/2}\cdot\hat{\bk} & \sqrt{2}\,T(\br) & 0 \\ \sqrt{2}\,T^\dagger(\br) & \hbar v_0\bm{\sigma}_{\theta/2}\cdot\hat{\bk} & U\,\sigma^0/2 \\  0 & U\,\sigma^0/2 & \hbar v_0\bm{\sigma}_{\theta/2}\cdot\hat{\bk}\end{array}\right),\quad
W=\frac{1}{\sqrt{2}}\left(\begin{array}{ccc}0 & \sigma^0 & \sigma^0 \\ \sqrt{2}\sigma^0 & 0 & 0 \\ 0 & \sigma^0 & -\sigma^0\end{array}\right),
\ee
which is the staring point in the main text.
In this basis we easily find that for $U=0$ the model \eqn{H_STTG_NB} is the sum of the TBG continuum model with interlayer hopping multiplied by $\sqrt{2}$ and an isolated Dirac cone. A finite displacement field, $U\neq 0$, breaks the $M_z$ symmetry and couples the additional Dirac cone with the bands of TBG.
Assuming a small twist angle $\theta$, we can neglect the phase factor in the Pauli matrices $\bm{\sigma}_{\pm\theta/2}\to\bm{\sigma}$ and the Hamiltonian \eqn{H_STTG} in the Moir\'e reciprocal lattice takes the form: 
\be
\label{H_STTG_RL}
H_{\bQ,\bQ^\prime}(\bk)=\hbar v_0\,\delta_{\bQ,\bQ^\prime}\,(\bk-\bQ)\cdot\bm{\sigma}+\frac{U}{2}\,\delta_{\bQ,\bQ^\prime}\,\sigma^0\,(l-2)+\sum_{j=1}^{3}\,T_j\left(\delta_{\bQ-\bQ^\prime,\bq_j}+\delta_{\bQ^\prime-\bQ,\bq_j}\right).
\ee
In the previous expression \eqn{H_STTG_RL} $\alpha$ and $\beta$ denote the sublattice indices ($\alpha,\beta=A,B$), MBZ stands for the Moir\'e Brillouin zone (grey shaded region in Fig. \ref{R_lattice_TTG}), $l=1,2,3$ denotes the bottom, middle and top layers, $\mathcal{Q}_{\pm}$ is the triangular lattice for layer $l=2$ and $l=1,3$, respectively, depicted as blue and red dots in Fig. \ref{R_lattice_TTG}. Finally, we denote with $\ket{u_{n\bk}}$ and $\ep_{n\bk}$ the eigenstates and eigenvalues of $H(\bk)$ \eqn{H_STTG_RL}. Notice that $H$ \eqn{H_STTG} has only three dimensionless parameter $\alpha=w/\hbar v_0\,k_\theta$, $r$ and $u=U/\hbar v_0\,k_\theta$.  
Before concluding the section we briefly list the symmetries of the Hamiltonian \eqn{H_STTG} that we are going to employ in the following. 
A detailed description of the symmetries of the model \eqn{H_STTG} is presented in Ref. \cite{calugaru2021tstg}.
At vanishing displacement field the Hamiltonian \eqn{H_STTG} commutes with the mirror symmetry $M_z$, the threefold rotation $C_{3z}$ around the $\bm{z}$ axis, the twofold rotation $C_{2x}$ around the in-plane $\bm{x}$ axis and the $C_{2z} T$ symmetry. 
In addition there are three useful transformations $P$, $C_{2x}\,P$ and $M_z\,C_{2x}\,P$. 
The first two, $P$ and $C_{2x}\,P$, act differently on the even sector $\{\ket{2},(\ket{1}+\ket{3})/\sqrt{2}\}$ (upper $2\times2$ block of the matrix in Eq. \eqn{H_STTG_NB}) and on the odd one $\{(\ket{1}-\ket{3})/\sqrt{2}\}$ (lower one-dimensional block of the matrix in Eq. \eqn{H_STTG_NB}). 
In particular we find that $P$ anticommutes with the projection of the Hamiltonian in the even sector, while $C_{2x}\,P$ anticommutes with both the even and odd mirror sectors. 
Moreover, $C_{2x}\,P$ commutes with the displacement field term coupling the even $(\ket{1}+\ket{3})/\sqrt{2}$ and the odd $(\ket{1}-\ket{3})/\sqrt{2}$ states. 
From the previous considerations, we readily realize that the anticommuting symmetry of the Hamiltonian at finite $U$ is obtained by combining $M_z$ and $C_{2x}\,P$ to form $M_z\,C_{2x}\,P$ ($\{M_z\,C_{2x}\,P,H\}=0$). The introduction of a displacement field breaks $M_z$ and $C_{2x}$ symmetries, while $C_{3z}$, $C_{2z}\,T$ and $M_z\,C_{2x}\,P$ remain proper symmetries of the Hamiltonian \eqn{H_STTG}. 

\subsection*{The evolution of the Dirac cones as a function of the displacement field}

\begin{figure}[ht]
\begin{subfigure}{.4\textwidth}
  \centering
  \includegraphics[width=0.9\linewidth]{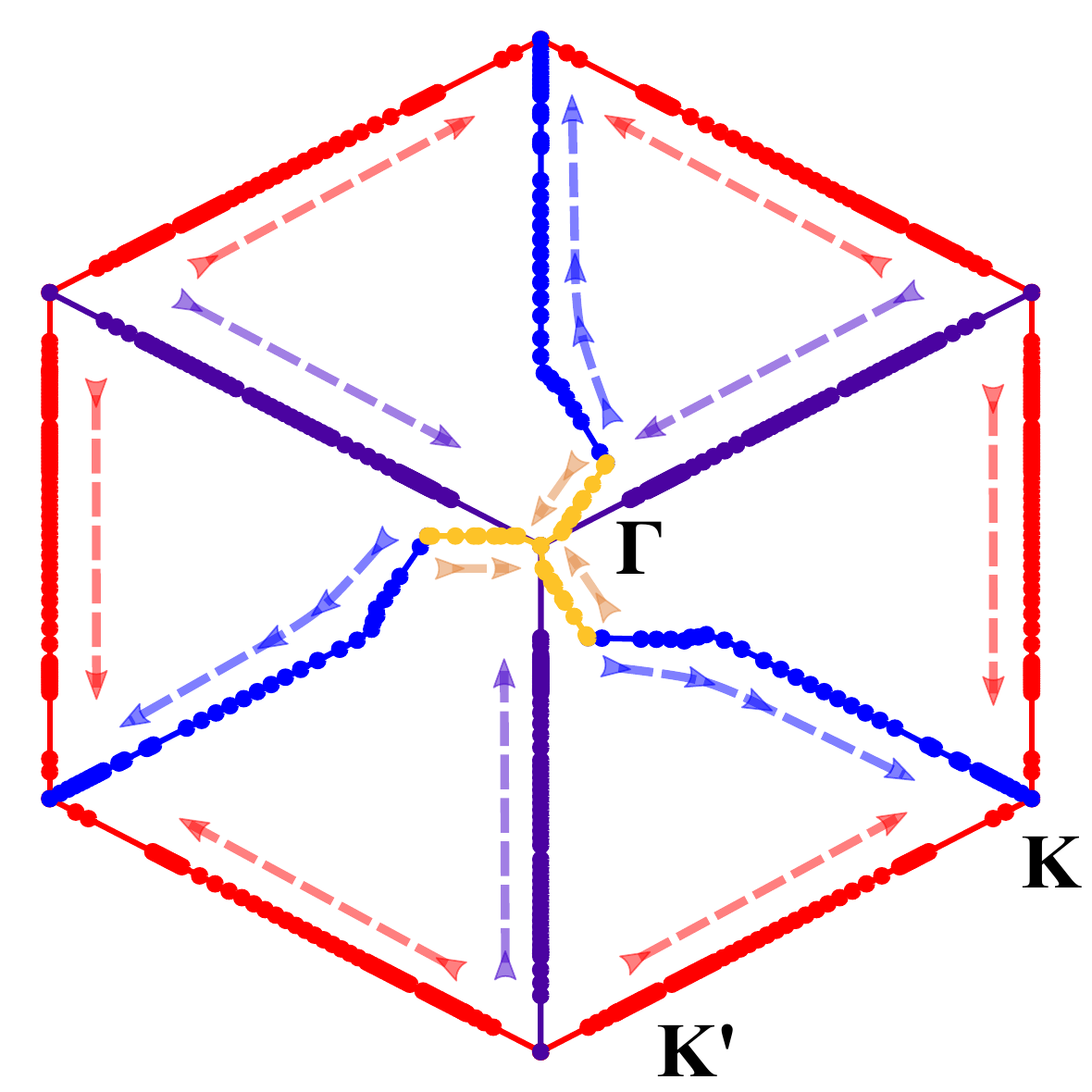}  
  \caption{Dirac cones evolution.}
  \label{DPs}
\end{subfigure}
\begin{subfigure}{.55\textwidth}
  \centering
  \includegraphics[width=0.9\linewidth]{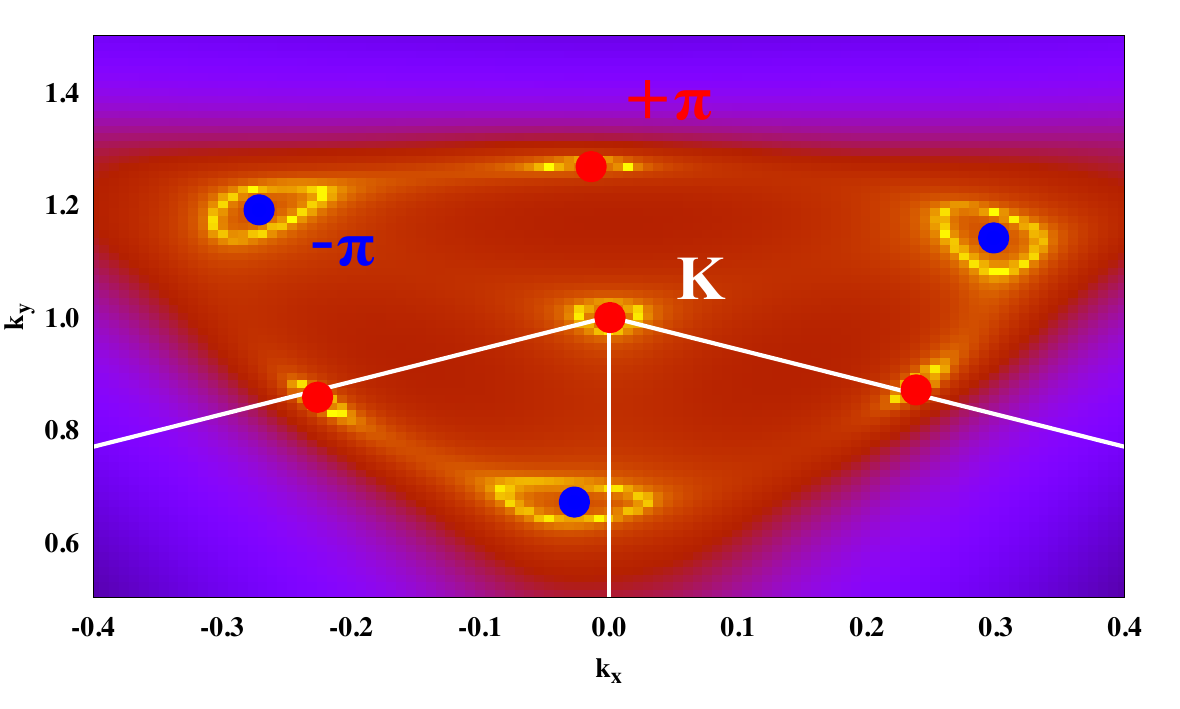}  
  \caption{momentum resolved density of states $\rho(\omega,\bk)$ around $\bK$.}
  \label{FS}
\end{subfigure}
\caption{(a) blue, red, violet and orange lines shows the evolution of the Dirac points. Red and violet Dirac points originate from $\bK^\prime$, by increasing $D$ the red data moves toward $\bK$ along the $\bK^\prime\bK$ line while violet data evolves toward $\bm{\Gamma}$ along the $\bK^\prime\bm{\Gamma}$ line. Orange and blue data are new Dirac points that are only present above $D/\ep_0\simeq 0.293 \text{V/nm}$. 
Finally, the red Dirac points merge with the Dirac point at $\bK$ for $D/\ep_0\simeq 0.366\text{V/nm}$ giving rise to the vanishing Fermi velocity. The calculation has been performed at $r=0.8$ and $\theta=1.59^\circ$. (b) Momentum resolved density of states around $\bK$ at $r=0.8$, $\theta=1.59^\circ$ and $D/\ep_0\simeq 0.34\text{V/nm}$. The red dots and blue dots denote Dirac points with Berry phase $+\pi$ (vorticity $+1$) and $-\pi$ (vorticity $-1$), respectively. }
\label{DPs_EVO}
\end{figure}

In this section we provide a simple picture for the generation of electric field induced higher-order VHS by the fusion of zero-energy Dirac cones. 
The evolution of the zero-energy Dirac cones in the MBZ as a function of the applied external electric displacement field is shown in Fig.~\ref{DPs_EVO}. The initial situation at zero displacement $D=0$ consists of a total of three zero-energy Dirac cones with the same vorticity $+1$: two located at $\bK^\prime$ and one at $\bK$. The Dirac cone at $\bK$ is protected by $C_{2z}\,T$ (forbids a gap opening) and  $M_z\,C_{2x}\,P$ (pins it at zero-energy by enforcing a local particle-hole symmetric spectrum). None of these symmetries is broken by the displacement field such there is always a Dirac cone at $\bK$ regardless of $D$. The situation is different at $\bK^\prime$ where the two zero-energy Dirac cones are shifted in momentum space as $D$ departs from zero, one along the $\bK^\prime\bK$ line (red dots) and the other one along the $\bK^\prime \bm{\Gamma}$ line (violet dots). Further increasing $D$, six new zero-energy Dirac cones appear at a critical $D\ge0.293\text{V/nm}$ in the middle of the Moiré Brillouin zone. There can be decomposed into two sets of threefold $C_{3 z}$-symmetric cones with opposite vorticities $\pm 1$ ($+1$ as orange dots, $-1$ as blue dots). The $+1$ Dirac cones further move towards $\bm{\Gamma}$ as $D$ increases, and the $-1$ Dirac cones towards $\bK$. Eventually, in the region close to the higher-order VHS where the velocity $v$ and the curvature $\eta$ are small ($D\sim0.34-0.366\,\text{V/nm}$ for $r=0.8$ and $\theta=1.59^\circ$), the landscape of Dirac cones close to $\bK$ is illustrated in Fig.~\ref{DPs_EVO}(b), with the symmetry-protected $+1$ cone at $\bK$ surrounded by six cones: three (red dots) of vorticity $+1$ originating from $\bK$ and three (blue dots) of opposite vorticity. The higher-order VHS thus corresponds exactly to the merging of all six side Dirac cones with the central one at the $\bK$ point. Above the critical value, $D>D_c$, only the isolated Dirac cone located $\bK$ with vorticity $+1$. 
\begin{figure}
\begin{center}
\includegraphics[width=0.8\textwidth]{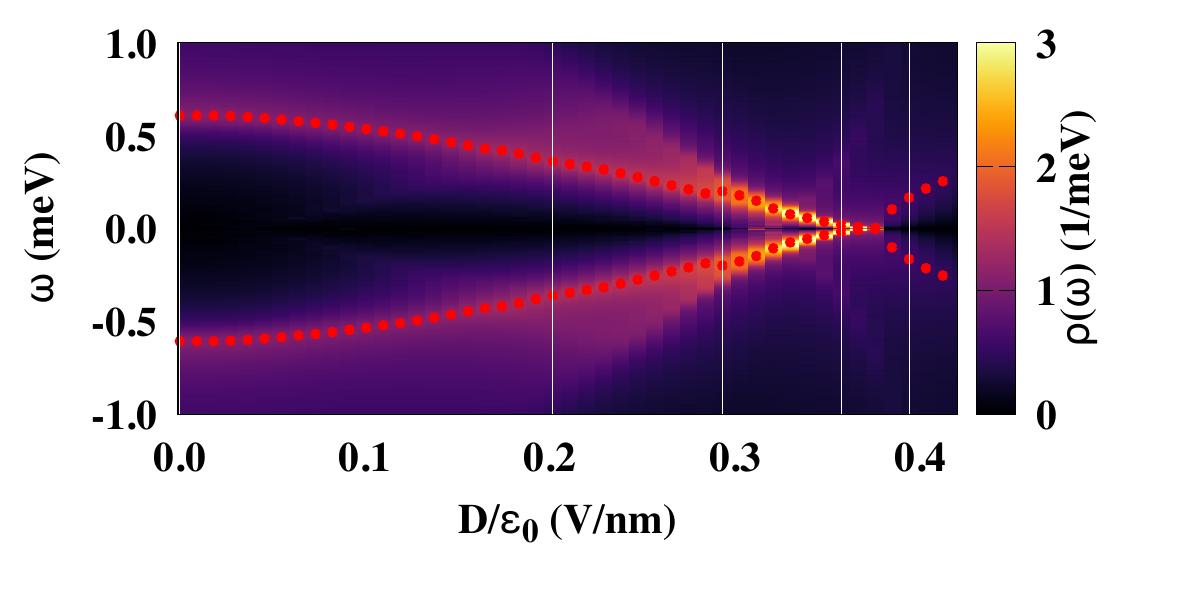}
\caption{evolution of the density of states as a function of the displacement and of the energy.}
\label{DOS_159_08}
\end{center}
\end{figure}
Finally, we report in Fig.~\ref{DOS_159_08} the evolution of the density of states (DOS) as a function of the energy and of the displacement field for $r=0.8$ and $\theta=1.59^\circ$. In correspondence of the point where the Fermi velocity vanishes $D/\ep_0\simeq 0.366\text{V/nm}$ we see a strong enhancement of the DOS close to charge neutrality.


\section{The $\bk\cdot\bp$ Hamiltonian at $\bK$}
\label{sec2}

The $\bk\cdot\bp$ Hamiltonian at $\bK$ is obtained by performing degenerate perturbation theory on the zero energy doublet $\{\ket{\Phi_\omega},\ket{\Phi_{\omega^*}}\}$ in the small perturbation $\delta H(\bq)=H(\bK+ \bq)-H(\bK)$ where $H(\bk)$ is given in Eq. \eqn{H_STTG_RL}, $\ket{\Phi_\omega}$ and $\ket{\Phi_{\omega^*}}$ are eigenstates of $H(\bK)$ and $C_{3z}$ with eigenvalues $\omega=e^{2\pi i/3}$ and $\omega^*$. In the following we derive the $\bk\cdot\bp$ Hamiltonian to third order in the small deviation $\bq=\bk-\bK$. 
To this aim we are going to show for each order in $\bq$ the terms allowed by the symmetries listed in the last part of Section \ref{sec1}.
Among them particularly important are $C_{3z}=e^{2\pi\tau^z i/3}$, $C_{2z} T=\tau^x\mathcal{K}$ with $\mathcal{K}$ complex conjugation and $M_z\,C_{2x}\,P=\tau^x$ that applied to the $\bk\cdot\bp$ Hamiltonian at $\bK$ give: 
\bal
\label{kp_symmetries}
&C_{3z}\,H_{\bK}(\bq)\,C^\dagger_{3z}=H_{\bK}(C_{3z}\,\bq),\\
&C_{2z}T\,H_{\bK}(\bq)\,\left(C_{2z}T\right)^\dagger=H_{\bK}(\bq),\\
&M_z\,C_{2x}\,P\,H_{\bK}(\bq)\,\left(M_z\,C_{2x}\,P\right)^\dagger=-H_{\bK}(-q_x,q_y).
\eal
At first order in perturbation theory we find that the only terms invariant under $C_{3z}$ are $q_-\,\tau^+$ and $q_+\,\tau^-$ where $q_{\pm}=q_x\pm iq_y$, $\tau^{\pm}=(\tau^x\pm i\tau^y)/2$, and $\tau^{0,x,y,z}$ are the Pauli matrices acting on the two-dimensional degenerate subspace $\{\ket{\Phi_\omega},\ket{\Phi_{\omega^*}}\}$. 
Moreover, any linear combination of the form $r\, q_+\,\tau^-+r^*\,q_-\,\tau^+$ with $r\in\mathbb{C}$ is invariant under the symmetry $C_{2z} T$. 
We employ the gauge freedom to choose the phases of the wave function $\ket{\Phi_{\omega}}$ and $\ket{\Phi_{\omega^*}}$ so that $r=\hbar v\in\mathbb{R}$. The latter constraint plays the role of a \textit{gauge-fixing} condition and gives the first order correction 
\be
H^{(1)}_{\bK}(\bq)=\hbar v\,\bq\cdot\bm{\tau}.
\ee 
The second order terms that are allowed by the $C_{3z}$ symmetry are $q^2_{-}\,\tau^-$, $q^2_{+}\,\tau^+$, and $q_-\,q_+\,\tau^0$. 
Let us start by looking at the first two contributions that give rise to the quadratic correction $f^*\,q^2_+\,\tau^++f\,q^2_-\,\tau^-=\text{Re}f\,\left[(q^2_x-q^2_y)\,\tau^x-2 q_x q_y\,\tau^y\right]+\text{Im}f\left[(q^2_x-q^2_y)\,\tau^y+2 q_x q_y\,\tau^x\right]$. From the particle-hole symmetry $M_{2z}\,C_{2x}\,P$ \eqn{kp_symmetries} it follows that $\text{Re}f=0$ and $f=-i\eta$. Additionally, as a consequence of $M_{2z}\,C_{2x}\,P$ the coefficient of $q_-\,q_+\,\tau^0$ vanishes. Finally, we find that the second order correction reads: 
\be
H^{(2)}_{\bK}(\bq)=\eta\,(q^2_y-q^2_x)\,\tau^y-2\eta q_x\,q_y\,\tau^x.
\ee
As a consequence of the $C_{3z}$ symmetry at third order in the wave vector $\bq$ we have $q^3_-\,\tau^0$, $q^3_+\,\tau^0$, $q^2_-q_+\,\tau^+$ and $q^2_+q_-\,\tau^-$. The latter two terms give rise to the cubic contribution $g\, q^2_+q_-\,\tau^-+g^*\,q^2_-q_+\,\tau^+=\text{Re}g\,q^2\left(\bq\cdot\bm{\tau}\right)+\text{Im}g\,q^2\left(q_x\,\tau^y-q_y\,\tau^x\right)$, that is invariant under $C_{2z}\,T$. Following the same line of reasoning used for the quadratic term, only $q^2\left(\bq\cdot\bm{\tau}\right)$ with $q^2=q^2_x+q^2_y$ is allowed by the particle-hole symmetry $M_z\,C_{2x}\,P$. We conclude that $\text{Im}g=0$ and $g=\xi\in\mathbb{R}$. Moreover, the same $M_z\,C_{2x}\,P$ symmetry selects the combination $\gamma(q^3_-+q^3_+)\,\tau^0$. Therefore, the third order correction reads: 
\be
H^{(3)}_{\bK}(\bq)=2\gamma(q^3_x-3q^2_y\,q_x)\,\tau^0+\xi q^2\left(\bq\cdot\bm{\tau}\right).
\ee
\begin{figure}
\begin{center}
\includegraphics[width=0.6\textwidth]{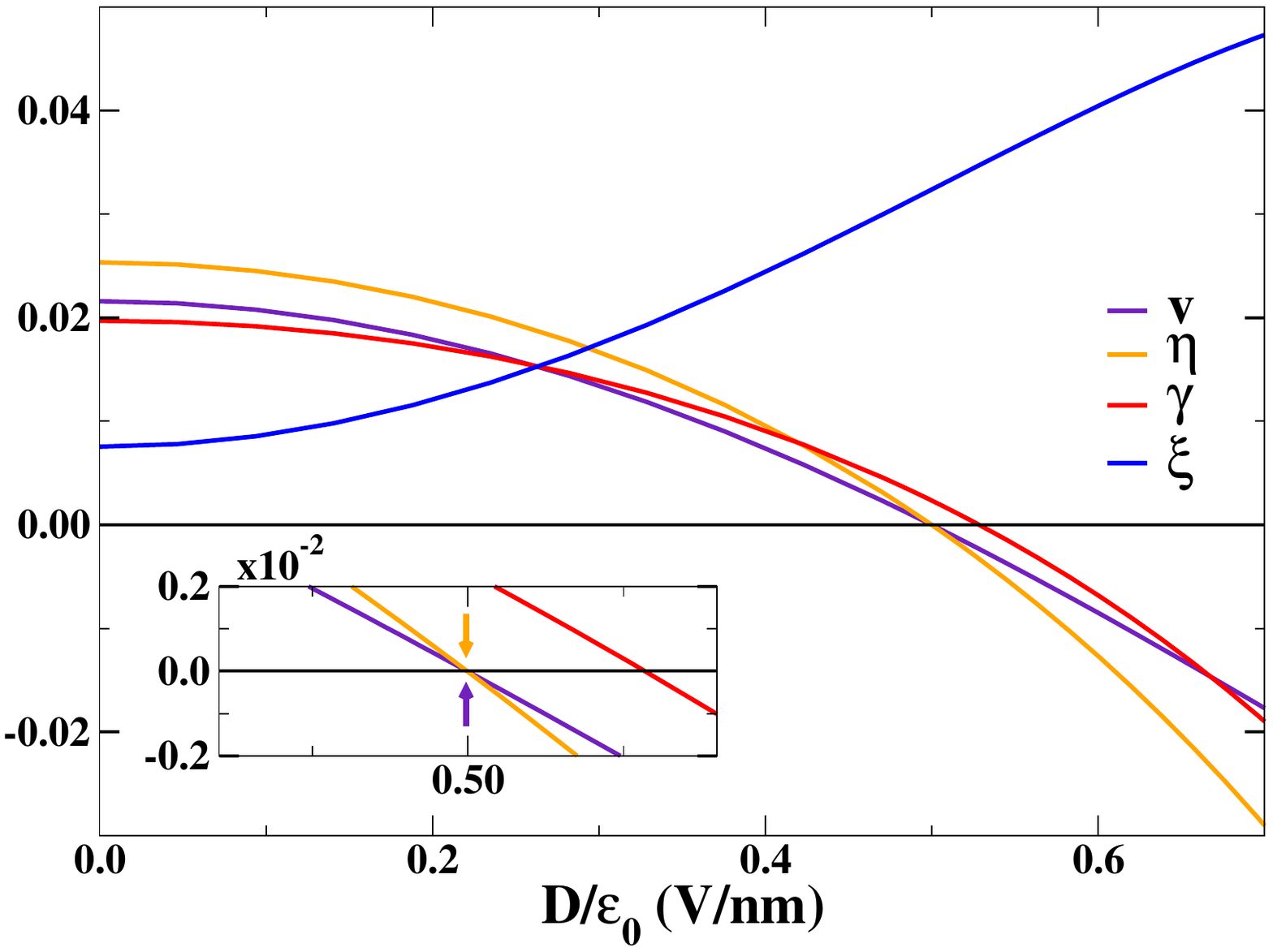}
\caption{evolution of the parameters of the $\bk\cdot\bp$ Hamiltonian \eqn{kp_H} at $\bK$ as a function of $D/\ep_0$ for $\theta=1.631^\circ$, $r=0.8$. The inset shows the region around the critical value $D_c$ where the parameters $v$ and $\eta$ vanish. }
\label{parameters_r08}
\end{center}
\end{figure}
By collecting the contributions the final Hamiltonian is, as given in the main text,  
\be
\label{kp_H}
H_{\bK}(\bq)=\hbar v\,\bq\cdot\bm{\tau}+\eta\,(q^2_y-q^2_x)\,\tau^y-2\eta q_x\,q_y\,\tau^x+2\gamma(q^3_x-3q^2_y\,q_x)\,\tau^0+\xi q^2\left(\bq\cdot\bm{\tau}\right)+O(q^4),
\ee
where the coefficients $v$, $\eta$, $\gamma$ and $\xi$ are obtained by computing the expectation values:
\begin{equation}
\label{parameters}
\begin{aligned}
&v=\bra{\Phi_{\omega^*}}\Sigma^-\ket{\Phi_\omega},\\
&\eta=-\text{Im}\left[\bra{\Phi_{\omega^*}}\Sigma^+\,\mathcal{P}\,\Sigma^+\ket{\Phi_{\omega}}\right],\\
&\gamma=\text{Re}\left[\bra{\Phi_{\omega}}\Sigma^-\,\mathcal{P}\,\Sigma^-\,\mathcal{P}\,\Sigma^-\ket{\Phi_{\omega}}\right],\\
&\xi=\text{Re}\left[\bra{\Phi_{\omega^*}}\Sigma^+\,\mathcal{P}\,\Sigma^-\,\mathcal{P}\,\Sigma^-+\Sigma^-\,\mathcal{P}\,\Sigma^+\,\mathcal{P}\,\Sigma^-+\Sigma^-\,\mathcal{P}\,\Sigma^-\,\mathcal{P}\,\Sigma^+\ket{\Phi_{\omega}}\right],
\end{aligned}
\end{equation}
 where $\Sigma^-_{\bQ\alpha,\bQ^\prime\beta}=\sigma^-_{\alpha\beta}\,\delta_{\bQ,\bQ^\prime}$ and we have introduced the projector in the high-energy states 
\begin{equation}
\label{projector_HE}
\mathcal{P}=-\sum_{n\notin L}\frac{\ket{u_n}\bra{u_n}}{\epsilon_n},     
\end{equation}
with $L=\{\ket{\Phi_\omega},\ket{\Phi_{\omega^*}}\}$, $\ket{u_n}=\ket{u_{n\bK}}$ and $\epsilon_n=\epsilon_{n\bK}$. 
Before going on we remark that the specific form of the $\bk\cdot\bp$ model at $\bK$~\eqn{kp_H} is determined by the $C_{3z}$ and the $M_z\,C_{2x}\,P$ symmetries. 
The breaking of one of these two symmetries would introduce additional $\bq$ dependent corrections to $H_{\bK}(\bq)$ that spoil the higher-order VHS at $\theta_c$ and $D_c$. Indeed, in the presence of additional quadratic corrections it is not possible to induce an higher-order VHS with the only two tuning parameters $\theta$ and $D$. 

\subsection{Properties of the higher-order VHS}

\begin{figure}
\begin{center}
\includegraphics[width=0.6\textwidth]{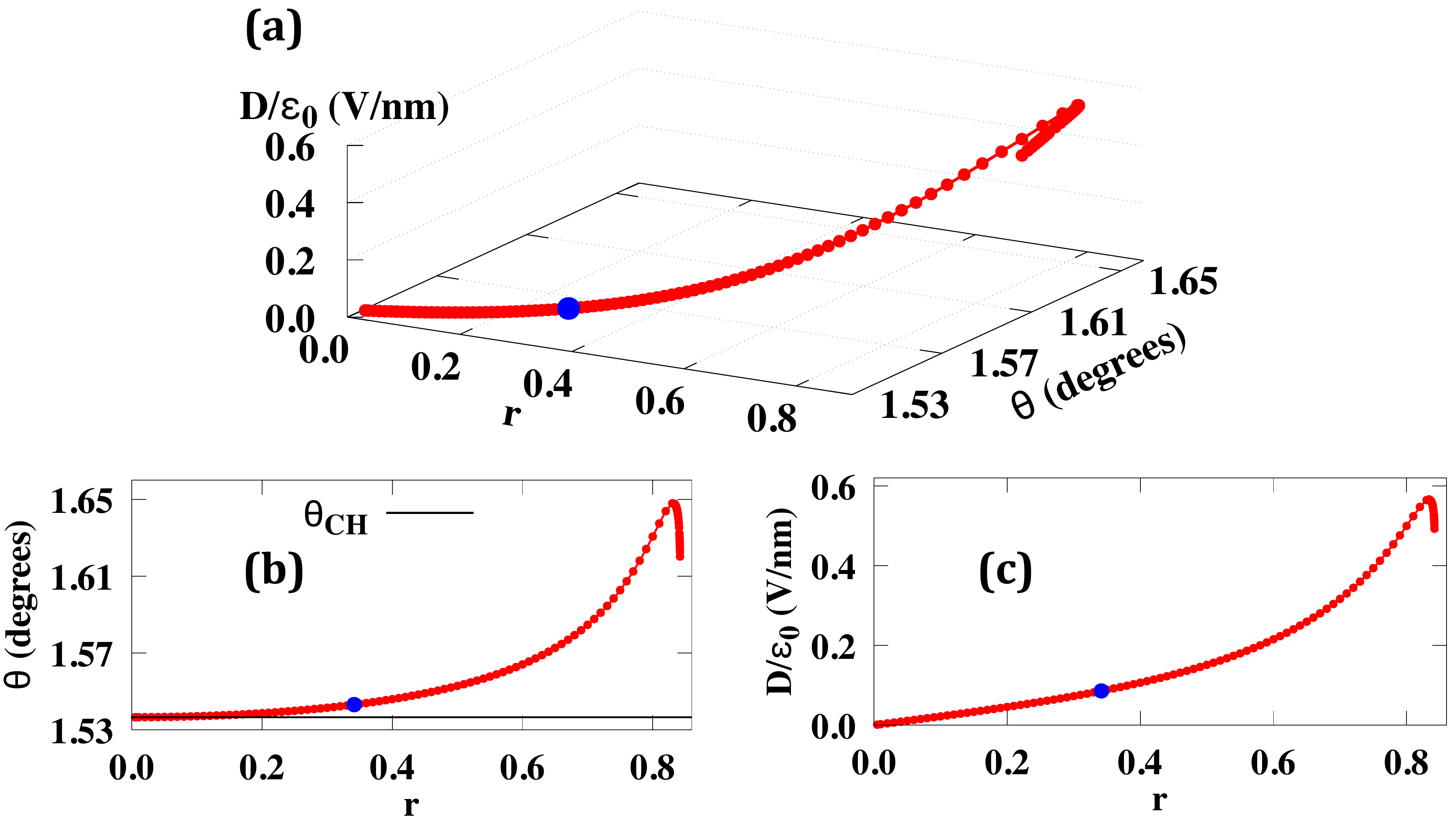}
\caption{(a) Three-dimensional trajectory of higher-order VHS originating from the first magic angle at $r=0$. (b-c) Projections along the $(r,\theta)$ and $(r,D)$ planes. The blue dot gives the position of the Lifshitz transition with anomalous exponent $\nu=-2/5$.}
\label{3D_HOVHS_1st}
\end{center}
\end{figure}
\begin{figure}
\begin{center}
\includegraphics[width=0.6\textwidth]{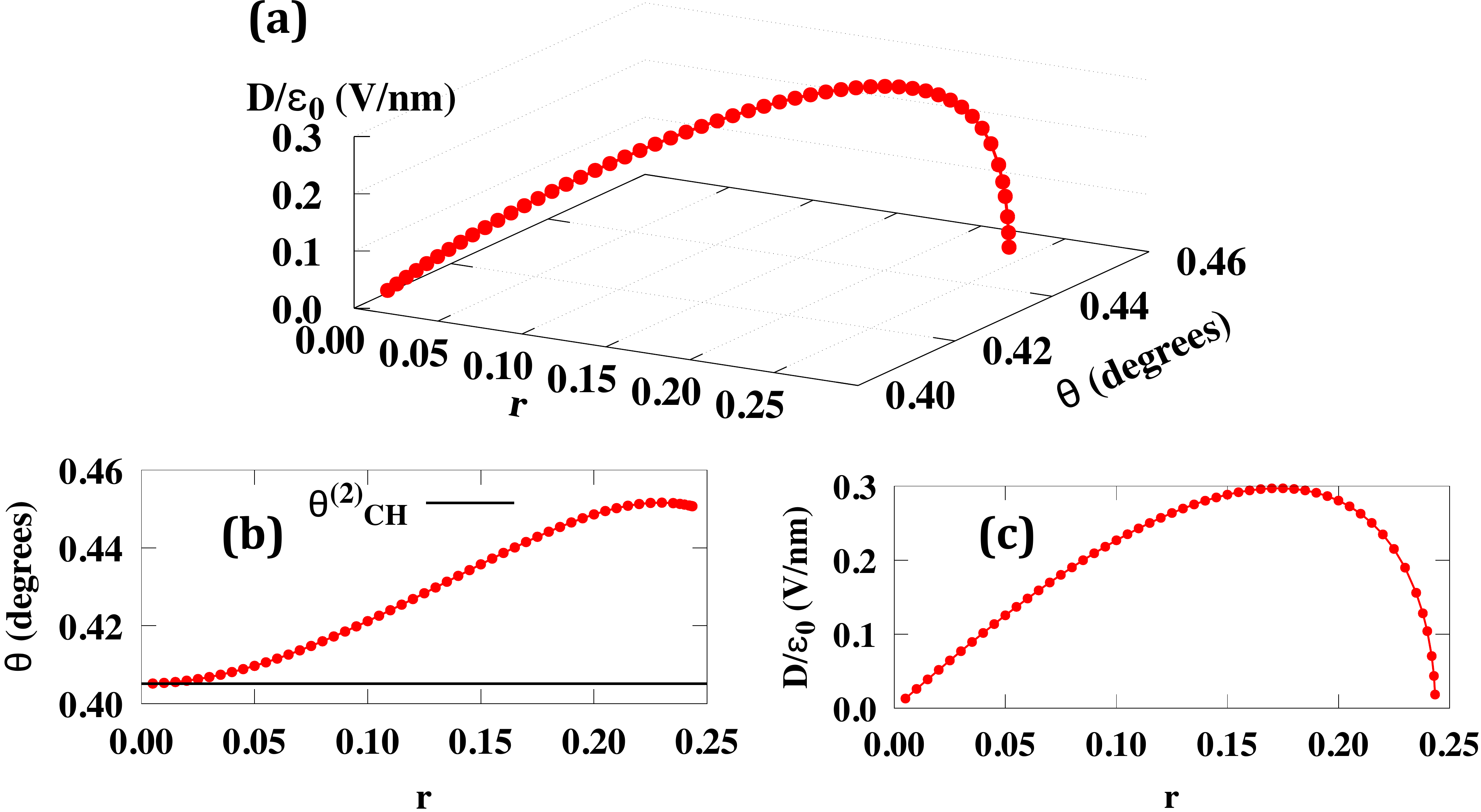}
\caption{ (a) Three-dimensional trajectory of higher-order VHS originating from the second magic angle at $r=0$. (b-c) Projections along the $(r,\theta)$ and $(r,D)$ planes.}
\label{3D_HOVHS_2nd}
\end{center}
\end{figure}

The evolution of the coefficients $\{v,\xi,\eta,\gamma\}$ for the critical value of the twist-angle $\theta_c\simeq1.631^\circ$ as a function of the displacement field $D$ at $r=0.8$ is shown in Fig. \ref{parameters_r08}. We find that when $D/\ep_0=D_c/\ep_0\simeq 0.5\,\text{V/nm}$ the $\bK$ point is an higher-order VHS where the first and second order coefficients both vanish $v=\eta=0$ as highlighted by the inset of Fig.~\ref{parameters_r08}.
Fig.~\ref{3D_HOVHS_1st} shows the trajectory of points where the higher-order VHS takes place together with its projections on the $(r,\theta)$ and $(r,D)$ planes.
As $D$ increase, the trajectory extends till the end point at $r\simeq0.842$ above which it is not possible to have $v$ and $\eta$ both vanishing. 
In the chiral limit $r=0$ the trajectory originates from the first magic angle $\theta_{CH}\simeq 1.536^\circ$ at vanishing displacement field. 
By reducing the twist-angle we find an additional trajectory of higher-order VHS, see Fig.~\ref{3D_HOVHS_2nd}, which starts from the second magic angle $\theta^{(2)}_{CH}\simeq 0.405^\circ$ at $r=0$ and $D=0$. In this case the end point of the trajectory is located at $r\simeq0.243$.
\begin{figure}[ht]
\begin{subfigure}{.325\textwidth}
  \centering
  \includegraphics[width=1.\linewidth]{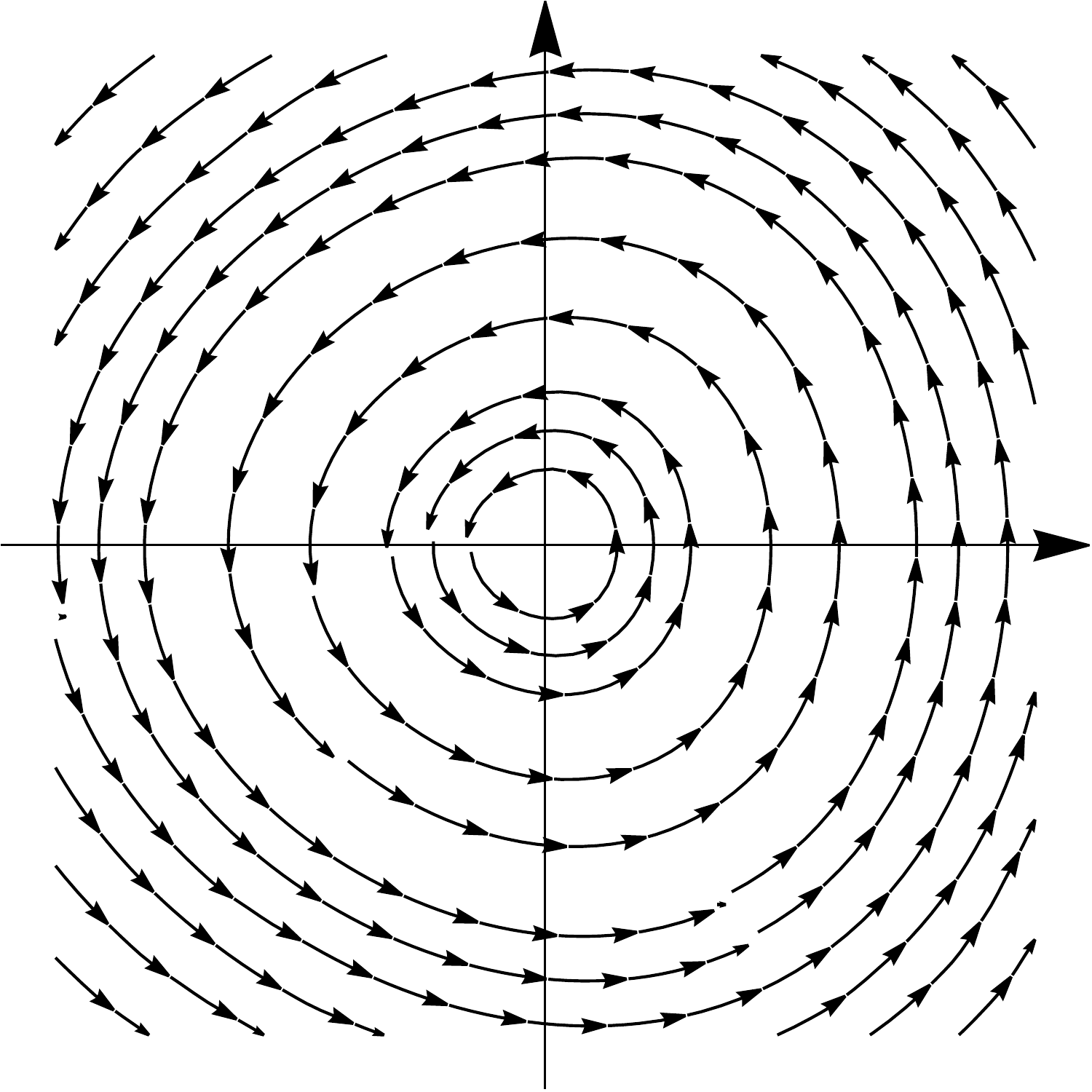}  
  \caption{$2\gamma/\xi=0.1$.}
  \label{energy_contours_1}
\end{subfigure}
\begin{subfigure}{.325\textwidth}
  \centering
  \includegraphics[width=1.\linewidth]{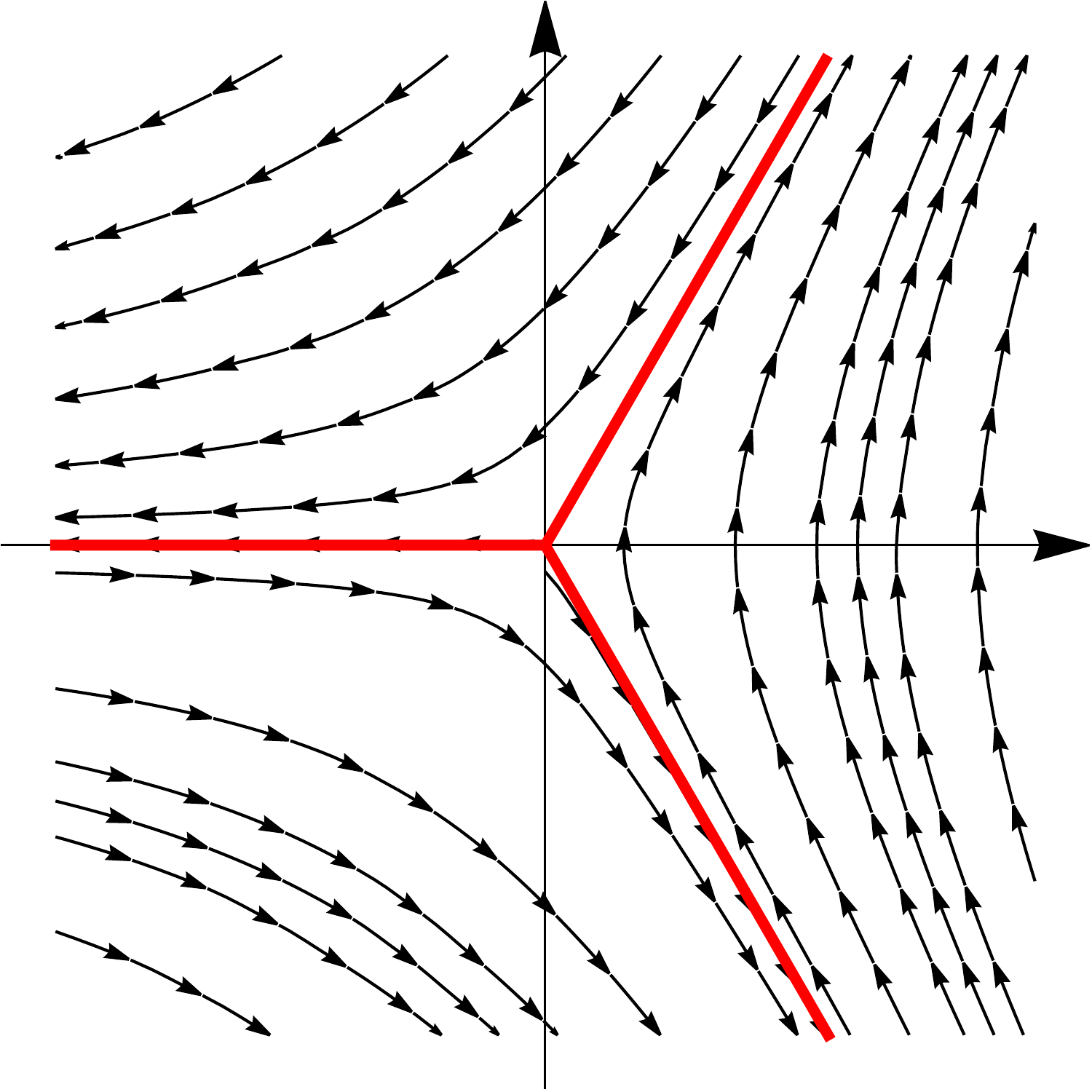}  
  \caption{$2\gamma/\xi=1.0$.}
  \label{energy_contours_2}
\end{subfigure}
\begin{subfigure}{.325\textwidth}
  \centering
  \includegraphics[width=1.\linewidth]{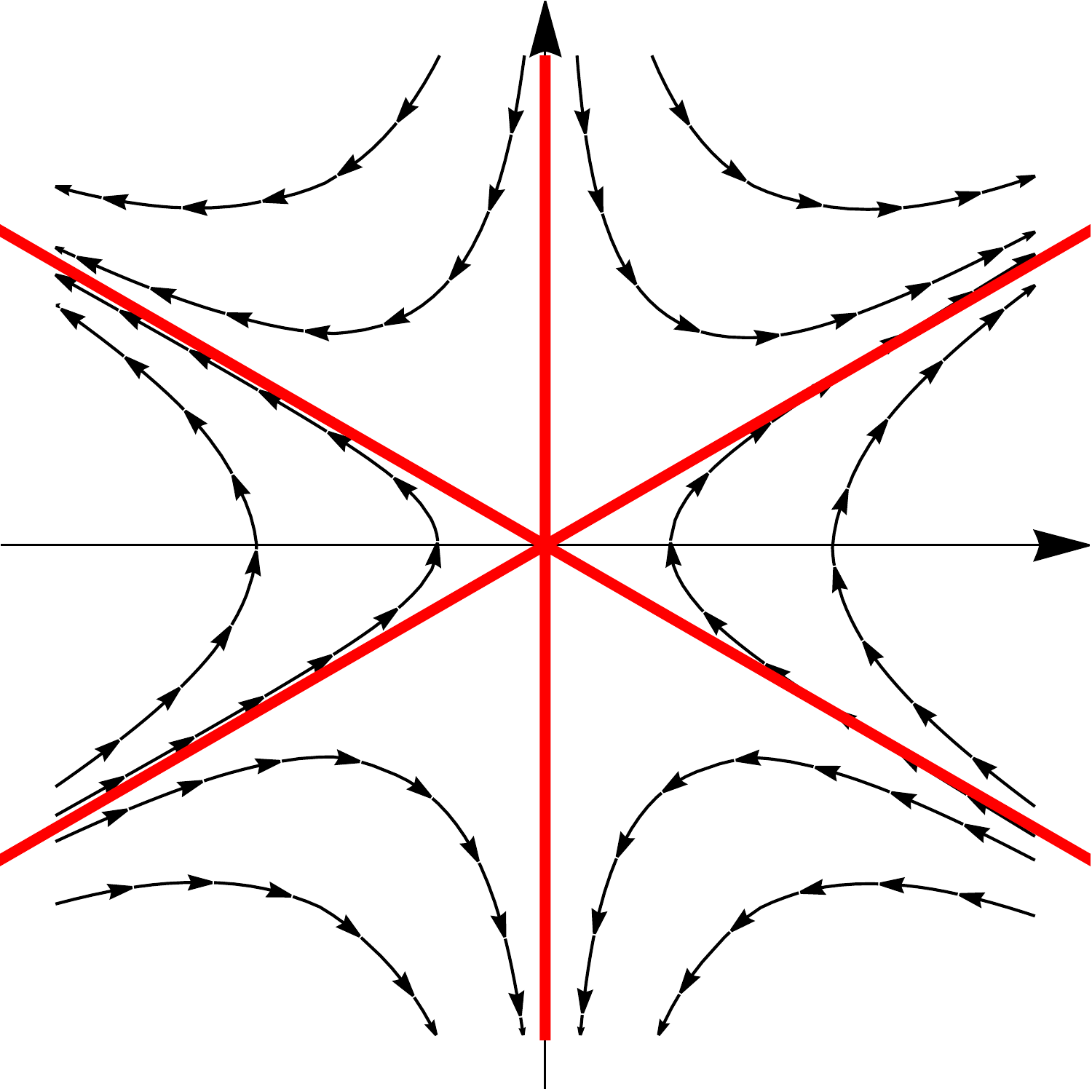}  
  \caption{$2\gamma/\xi=20.0$.}
  \label{energy_contours_3}
\end{subfigure}
\caption{constant energy lines of positive eigenvalue $E_+(\bq)$ of $H_{\bK}(\bq)$ at the higher-order VHS for three different values of $2\gamma/\xi$. Red solid lines show the asymptotes of the open trajectories. The energy contours of $E_-(\bq)$ are obtained by employing the particle-hole symmetry $M_z\,C_{2x}\,P$.}
\label{energy_contours}
\end{figure}

By assuming that $\xi,\gamma>0$ near the higher-order VHS the density of states reads: 
\bal
\label{DOS}
\rho(\omega)&=\sum_{s=\pm}\int \,\frac{d^2\bq}{4\pi^2}\,\delta\left[\omega-E_{s}(\bq)\right]=\frac{|\omega|^{-\frac{1}{3}}}{12\pi^2\,\xi^{2/3}}\Bigg[\int_{0}^{\phi_-}\frac{d\phi}{\left[1+(2\gamma/\xi)\cos\phi\right]^{2/3}}+\int_{\phi_+}^{\pi}\frac{d\phi}{\left[1-(2\gamma/\xi)\cos\phi\right]^{2/3}}\\
&+\int_{\phi_-}^{\pi}\frac{d\phi}{\left[-1-(2\gamma/\xi)\cos\phi\right]^{2/3}}+\int_{0}^{\phi_+}\frac{d\phi}{\left[(2\gamma/\xi)\cos\phi-1\right]^{2/3}}\Bigg]=\frac{|\omega|^{-\frac{1}{3}}}{\xi^{2/3}}\mathcal{F}\left(\frac{2\gamma}{\xi}\right),
\eal
where $\phi_+=\arccos(\xi/2\gamma)$ if $\xi/2\gamma\le1$ else $\phi_+=0$, while $\phi_-=\arccos(-\xi/2\gamma)$ if $\xi/2\gamma\le1$ else $\phi_-=\pi$.
We find the DOS is power law divergent with $\nu=-1/3$ and particle-hole symmetric. Interestingly, at the higher-order VHS there are two parameters $\gamma$ and $\xi$ that determines the properties of the iso-energy lines around $\bK$. 
The qualitative evolution of the energy contours as a function of the only dimensionless parameter $2\gamma/\xi$ at the higher-order VHS ($v=\eta=0$) is shown in Fig. \ref{energy_contours}. We observe that in the regime of $2\gamma/\xi\ll1$, see Fig. \ref{energy_contours_1} calculated at $2\gamma/\xi=0.1$, the iso-energy lines are closed orbits almost rotationally symmetric winding around $\bK$. By taking the limit $\gamma\to0$ in Eq. \eqn{DOS} the DOS reads $\rho(\omega)=|\omega|^{-1/3}/(6\pi\xi^{2/3})$. 
As we increase the ratio $2\gamma/\xi$ the iso-energy lines deform and the $U(1)$ rotational symmetry is lowered to the discrete $C_{3z}$ symmetry. 
Interestingly, by further increasing the ratio $2\gamma/\xi$ the effective model \eqn{kp_H} shows a Lifshitz transition at $2\gamma/\xi=1$ where the iso-energy lines are open hyperbolic orbits with asymptotes forming an angle $\pi/3+(j-1)2\pi/3$, $j=1,2,3$, with the $x$ axis.
Correspondingly, as shown in the main text the prefactor $\mathcal{F}\left(2\gamma/\xi\right)$ defined in Eq. \eqn{DOS} diverges and the power-law singularity becomes stronger than $-1/3$. In the next section we will show that higher-order corrections fix to $\nu=-2/5$ the power-law singularity at the Lifshitz transition.
Finally, for $2\gamma/\xi\gg1$, see Fig.~\ref{energy_contours_3}, the energy contours are hyperbolic orbits characterized by asymptotes $\pi/6+(j-1)\pi/3$, $j=1,2,3,4,5,6$, with the $x$ axis. These iso-energy contours have also been predicted in ABC stacking trilayer graphene on boron nitride in Ref. \cite{Fu_2019} and in biased bilayer graphene in Ref. \cite{PhysRevB.95.035137}. By taking the limit $\xi/2\gamma\to0$ in Eq. \eqn{DOS} we obtain $\rho(\omega)=\Gamma(7/6)|\omega|^{-1/3}/[\pi^{3/2}(2\gamma)^{2/3}\Gamma(2/3)]$ where $\Gamma(z)$ is the Euler Gamma function.
The evolution of $\xi$ and $2\gamma$ along the trajectory of higher-order VHS originating from the first magic angle at $r=0$ is shown in the main text. 
Starting from $r\simeq0.81$ we find at first that the value of $\xi$ is much larger that $\gamma$, $2\gamma/\xi\ll1$. 
Then, by reducing the atomic corrugation $r$ the ratio $2\gamma/\xi$ displayed in the main text grows monotonically till the value $r\simeq0.34$ ($\theta_c\simeq1.543^\circ$, $D/\ep_0\simeq0.086\,\text{V/nm}$) where $2\gamma/\xi$ crosses the critical point 1. 
Here, the Lifshitz transition takes place and the power-law singularity of the density of states changes from $\nu=-1/3$ to $-2/5$. 
Finally, for $r<0.34$, $2\gamma/\xi$ becomes larger than 1 and increases approaching the chiral limit $r=0$. 
Thus, by changing the value of $r$ the constant energy lines at the higher-order VHS evolves from $2\gamma/\xi\ll1$ to $2\gamma/\xi\gg1$.

\subsection{The density of states at the Lifshitz transition} 

At the Lifshitz transition $2\gamma=\xi$ and to leading order in the $\bk\cdot\bp$ expansion the dispersion relation reads: 
 \be
 E_{\pm}(\bq)=\xi q^3 (\pm1 + \cos3\phi),
 \ee
 where $q=\sqrt{q^2_x+q^2_y}\ge0$.
The vanishing of the dispersion relation $E_{+}(\bq)$ and $E_{-}(\bq)$ along the lines $\phi^{+}_{j}=\pi/3+2\pi(j-1)/3$ and $\phi^{-}_{j}=2\pi(j-1)/3$ with $j=1,2,3$ gives rise to the divergence of $\mathcal{F}(2\gamma/\xi)$ in Eq. \eqn{DOS} as displayed in the main text. In order to characterize the singularity we include next to leading order corrections in the wave vector $\bq$ to the $\bk\cdot\bp$ Hamiltonian \eqn{kp_H}.
By employing the symmetries of the model \eqn{kp_symmetries} at $\bK$ we find that the $q^4$ correction reads: 
\be
H^{(4)}_{\bK}(\bq)=\left[-2\nu\,q_x\,q_y\,q^2-4\mu q_x\,q_y(q^2_x-q^2_y)\right]\,\tau^x+\left[\nu\,(q^4_x-q^4_y)+\mu\,(q^4_x-6\,q^2_x\,q^2_y+q^4_y)\right]\,\tau^y,
\ee
while the $q^5$ one is: 
\be
H^{(5)}_{\bK}(\bq)=2\zeta (q^3_x-3q^2_y\,q_x)\,q^2\,\tau^0+\Upsilon\,q^4\,(\bq\cdot\bm{\tau})+w\left[\tau^x(q^5_x-10q^3_x\,q^2_y+5q_x\,q^4_y)+\tau^y(-q^5_y+10q^3_y\,q^2_x-5q^4_x\,q_y)\right].
\ee 
By taking into account $H^{(4)}_{\bK}(\bq)$ and $H^{(5)}_{\bK}(\bq)$, and expanding to the fifth order the dispersion relation, we find: 
\be
E_{\pm}(\bq)=\xi q^3 (\pm1 + \cos3\phi)\mp(\mu+\nu)\,q^4\,\sin3\phi+\left[2\zeta\cos3\phi\pm\frac{(\mu-\nu)^2+4w \xi+(4\xi\Upsilon+(\mu-\nu)^2)\cos6\phi}{4\xi}\right]\,q^5.
\ee  
From the latter expression, we realize that the fourth order term, proportional to $\propto q^4\,\sin3\phi$, vanishes along the directions $\phi^+_{j}$ and $\phi^-_{j}$ where the singularity in the angular integral takes place and the first non-vanishing correction is $q^5$. 
\begin{figure}
\begin{center}
\includegraphics[width=0.6\textwidth]{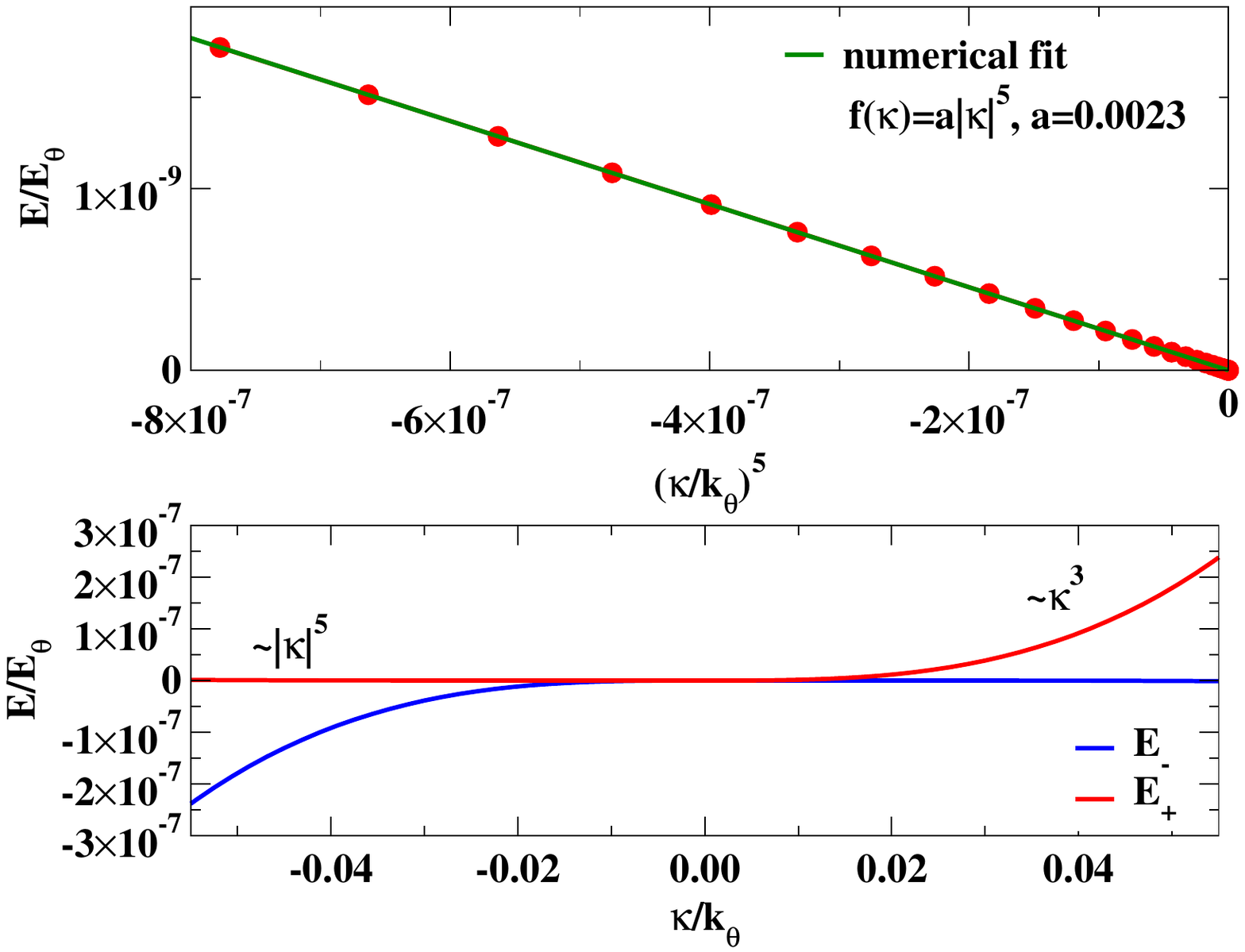}
\caption{top panel: red data shows the dispersion relation $\ep_{+\bK+\kappa\bm{x}}$ obtained by full numerical diagonalization of the Hamiltonian \eqn{H_STTG} as a function of $(\kappa/k_\theta)^5$ for $\kappa<0$. Green solid line is the fit of the data with the function $a|\kappa|^5$ where $\Lambda= a\simeq0.0023$. Bottom panel: red and blue lines show $\ep_{+\bK+\kappa\bm{x}}$ and $\ep_{-\bK+\kappa\bm{x}}$, respectively, for $\kappa/k_\theta\in[-0.06,0.06]$. The results are obtained by performing full numerical diagonalization of the model \eqn{H_STTG} at the higher-order VHS where the Lifshitz transition takes place $r\simeq0.34$, $\theta\simeq1.543^\circ$ and $D/\ep_0\simeq0.086\,\text{V/nm}$. The energy is measured in unit of $E_\theta=\hbar v_0 k_\theta$.}
\label{q_dependence_HOVHS_LT}
\end{center}
\end{figure}
The bottom panel of Fig. \ref{q_dependence_HOVHS_LT} shows the lowest energy electronic bands $\ep_{\pm\bk}$ along the line $\bk=\bK+\bm{\kappa}$ with $\bm{\kappa}\parallel k_x$ obtained by diagonalizing the model \eqn{H_STTG} at the higher-order VHS ($2\gamma=\xi$).  In agreement with the results of the $\bk\cdot\bp$ expansion the top panel of Fig. \ref{q_dependence_HOVHS_LT} shows that for $\kappa=-q<0$ the dispersion $\ep_{+\bK-q\bm{x}}$ goes as $\sim q^5$ while for positive wave vectors deviations $\kappa=q>0$ we find $\ep_{+ \bK+q\bm{x}}\sim q^3$. 

To determine the power-law singularity of the density of states we expand the dispersion relation $E_{\pm}(\bq)$ for angles $\phi$ close to the singular lines $\phi^{\pm}_j$ keeping terms up to the fifth order: 
\be
E_{\pm}\left(q,\phi^{\pm}_j+\frac{\delta\phi}{3}\right)\simeq\pm\frac{\xi}{2}\,\delta\phi^2\,q^3+(\mu+\nu)\,\delta\phi\,q^4\pm \Lambda q^5,\quad \Lambda=\frac{(\mu-\nu)^2+4w \xi+(4\xi\Upsilon+(\mu-\nu)^2)}{4\xi}-2\zeta,
\ee
where as shown in Fig. \ref{q_dependence_HOVHS_LT} the parameter $\Lambda$ is positive $\Lambda>0$.
By considering $\omega>0$ the wave vector $q_0$ that satisfies the delta function constraint in Eq. \eqn{DOS} is now solution of 
\be
\label{delta_function_root_eq}
\omega=\xi\,q^3\,\delta\phi^2/2+(\mu+\nu)\,\delta\phi\,q^4+\Lambda\,q^5.
\ee
In the regime $\delta\phi\ll1$ the first two terms in the left hand side are negligible and Eq. \eqn{delta_function_root_eq} is solved by $q_{0,<}(\delta\phi)=\left(\omega/\Lambda\right)^{1/5}$. On the other hand, in the opposite regime of $\delta\phi>1$ we find that the first contribution on the right hand side of Eq. \eqn{delta_function_root_eq} is the dominant term and $q_{0,>}(\delta\phi)=\left(2\omega/\xi\delta\phi^2\right)^{1/3}$. The middle term in Eq. \eqn{delta_function_root_eq} never plays a role in determining the asymptotic behaviour. We can estimate the location $\delta\phi_c$ of the crossover between the two regimes by matching the two asymptotic behaviours $\left(\omega/\Lambda\right)^{1/5}=\left(2\omega/\xi\delta\phi^2_c\right)^{1/3}$ that gives $\delta\phi_c=\Lambda^{3/10}\omega^{1/5}\sqrt{2/\xi}$.
By following the same line of reasoning we can split the angular integral in the density of states in two contributions corresponding to the regimes of small and large values of $\delta\phi$: 
\be
\rho(\omega)\propto\underbrace{\int_{0}^{\delta\phi_c} \frac{d\delta\phi}{5\Lambda\,q^3_{0,<}(\delta\phi)}}_{\propto \omega^{-2/5}}+\underbrace{\int_{\delta\phi_c}^{} \frac{d\delta\phi}{3\xi\,q_{0,>}(\delta\phi)\,\delta\phi^2}}_{\propto \omega^{-2/5}}\propto\omega^{-2/5}.
\ee
Thus, we conclude that at the Lifshitz transition at $2\gamma=\xi$ the density of states exhibits the energy scaling $\rho(\omega)\propto\omega^{\nu}$ with $\nu=-2/5$. As a consequence of the change in the Fermi surface topology, this exponent differs from the one $\nu=-1/3$ expected from simple scaling arguments.

\section{The Landau level spectrum}
\label{sec3}

In this section we derive the low-energy Hamiltonian that describes the Landau levels in the proximity of $\bK$.
We treat the magnetic field as a small perturbation which is introduced in the continuum model \eqn{H_STTG} by performing the minimal substitution $\bq\to-i\nabla_\br+e\bm{A}(\br)/\hbar$:
\be
\delta H_{\bQ,\bQ^\prime}(\hat{\bm{\pi}})=H_{\bQ,\bQ^\prime}(\bK+\hat{\bm{\pi}})-H_{\bQ,\bQ^\prime}(\bK)=\hbar v_0\,k_\theta\,\delta_{\bQ,\bQ^\prime}\,(\sigma^-\,\hat{\pi}_++\sigma^+\,\hat{\pi}_-)/(l_B k_\theta),
\ee 
where $l_B=\sqrt{\hbar/eB}$ is the magnetic length.
In the following we set $E_\theta=\hbar v_0\, k_\theta$ as the energy unit and $g=l_B\,k_\theta\propto 1/\sqrt{B}$.
In the small magnetic field limit $g\gg1$.
It is important to notice that $\hat{\pi}_+=(\hat{\pi}_x+i\hat{\pi}_y)\,l_B$ and $\hat{\pi}_-=(\hat{\pi}_x-i\hat{\pi}_y)\,l_B$ are now conjugate quantum operators. 
In particular, by choosing the Landau gauge where $\bm{A}(\br)=(0,Bx,0)$ we have $\hat{\pi}_x=\hat{q}_x$ and $\hat{\pi}_y= q_y+\hat{x}/l^2_B$ which give $[\hat{\pi}_y,\hat{\pi}_x]=i/l^2_B$.
By introducing the ladder operators $a$ and $a^\dagger$ so that $\hat{q}_x\,l_B=-i(a-a^\dagger)/\sqrt{2}$ and $\hat{x}/l_B=(a+a^\dagger)/\sqrt{2}$ ($\pi_-=-\sqrt{2} i a$, $\pi_+=\sqrt{2} i a^\dagger$) we find: 
\be
\delta H_{\bQ,\bQ^\prime}(\hat{\bm{\pi}})=\sqrt{2}\,i\,\delta_{\bQ,\bQ^\prime}\,(\sigma^-\,a^\dagger-\sigma^+\,a)/g.
\ee 
We are now ready to perform degenerate perturbation theory in the two-dimensional subspace $\left\{\ket{\Phi_\omega},\ket{\Phi_{\omega^*}}\right\}$.
Due to the $C_{3z}$ symmetry at first order in $\delta H(\hat{\bm{\pi}})$ only the matrix element $\bra{\Phi_{\omega^*}}\Sigma^-\ket{\Phi_\omega}$ and its complex conjugate are not vanishing. We find:
\be
H^{(1)}_{\bK}=\sqrt{2}i\,v\left(\tau^-\,a^\dagger-\tau^+\,a\right)/g,
\ee 
where the parameter $v$ is given in the first line of Eq. \eqn{parameters} and $\tau^+=\ket{\Phi_\omega}\bra{\Phi_{\omega^*}}$ and $\tau^-=\ket{\Phi_\omega^*}\bra{\Phi_{\omega}}$ are defined in the subspace $\left\{\ket{\Phi_\omega},\ket{\Phi_{\omega^*}}\right\}$.
The second order correction is given by: 
\be
H^{(2)}_{\bK}=\sum_{z,z^\prime}\ket{\Phi_z}\bra{\Phi_z}\delta H(\hat{\bm{\pi}})\,\mathcal{P}\,\delta H(\hat{\bm{\pi}})\ket{\Phi_{z^\prime}}\bra{\Phi_{z^\prime}},
\ee 
where $\mathcal{P}$ is the projector defined in Eq. \eqn{projector_HE}. As a consequence of the $M_z\,C_{2x}\,P$ symmetry we have $\bra{\Phi_z}\,\Sigma^-\,\mathcal{P}\,\Sigma^+\ket{\Phi_z}=0$, and $\bra{\Phi_{\omega^*}}\,\Sigma^+\,\mathcal{P}\,\Sigma^+\ket{\Phi_{\omega}}=-i\eta,\quad \eta\in\mathbb{R}$ in agreement with the second line of Eq. \eqn{parameters}. Thus, we find
\be
H^{(2)}_{\bK}=2i\eta\,\left(\tau^-\,a\,a-\tau^+\,a^\dagger\,a^\dagger\right)/g^2.
\ee 
The third order correction is given by: 
\be
H^{(3)}_{\bK}=\sum_{z,z^\prime}\ket{\Phi_z}\bra{\Phi_z}\delta H(\hat{\bm{\pi}})\,\mathcal{P}\,\delta H(\hat{\bm{\pi}})\,\mathcal{P}\,\delta H(\hat{\bm{\pi}})\ket{\Phi_{z^\prime}}\bra{\Phi_{z^\prime}}.
\ee 
Due to the $M_z\,C_{2x}\,P$ symmetry we have:
\be
\gamma=\bra{\Phi_z}\Sigma^-\,\mathcal{P}\,\Sigma^-\,\mathcal{P}\,\Sigma^-\ket{\Phi_z}=\bra{\Phi_z}\Sigma^+\,\mathcal{P}\,\Sigma^+\,\mathcal{P}\,\Sigma^+\ket{\Phi_z},\quad\gamma\in\mathbb{R}, 
\ee
and 
\be
b=\bra{\Phi_{\omega^*}}\,\Sigma^-\,\mathcal{P}\,\Sigma^-\,\mathcal{P}\,\Sigma^+\ket{\Phi_{\omega}}=\bra{\Phi_{\omega^*}}\,\Sigma^+\,\mathcal{P}\,\Sigma^-\,\mathcal{P}\,\Sigma^-\ket{\Phi_{\omega}}, \quad b\in\mathbb{R},
\ee
\be
c=\bra{\Phi_{\omega^*}}\,\Sigma^-\,\mathcal{P}\,\Sigma^+\,\mathcal{P}\,\Sigma^-\ket{\Phi_{\omega}}, \quad c\in\mathbb{R},
\ee
where $c+2b=\xi$ and $\xi$ is given in the last line of Eq. \eqn{parameters}.
After some algebra we find: 
\be
\label{3rd}
H^{(3)}_{\bK}=2i\sqrt{2}\gamma\,\tau^0\,(a\,a\,a-a^\dagger\,a^\dagger\,a^\dagger)/g^3+2i\sqrt{2}\xi(\tau^-\,a^\dagger\,a\,a^\dagger-\tau^+\,a\,a^\dagger\,a)/g^3.
\ee
\begin{figure}[ht]
\begin{subfigure}{.49\textwidth}
  \centering
  \includegraphics[width=0.9\linewidth]{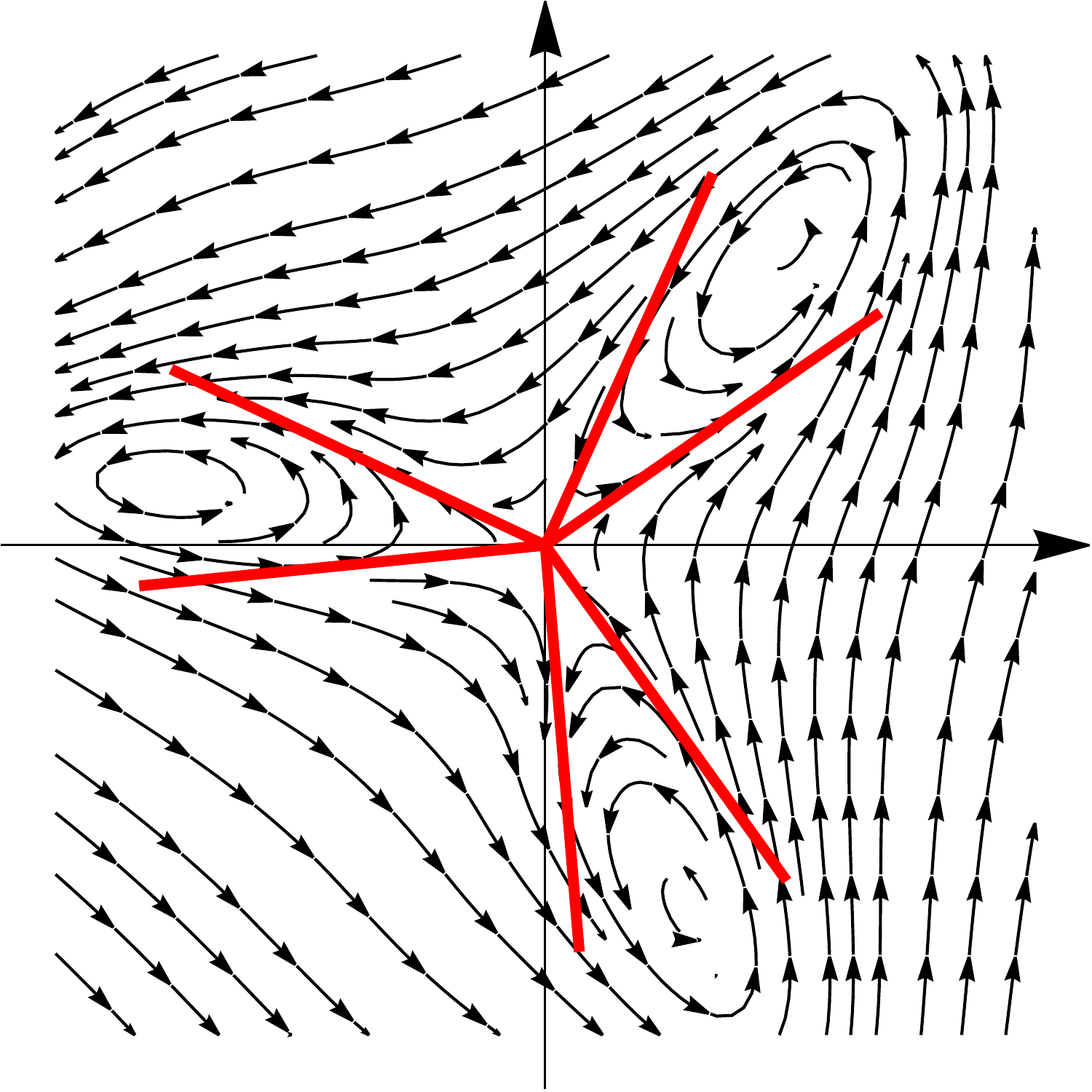}  
  \caption{$2\gamma/\xi=1.73$.}
  \label{energy_contours_q4_A}
\end{subfigure}
\begin{subfigure}{.49\textwidth}
  \centering
  \includegraphics[width=0.9\linewidth]{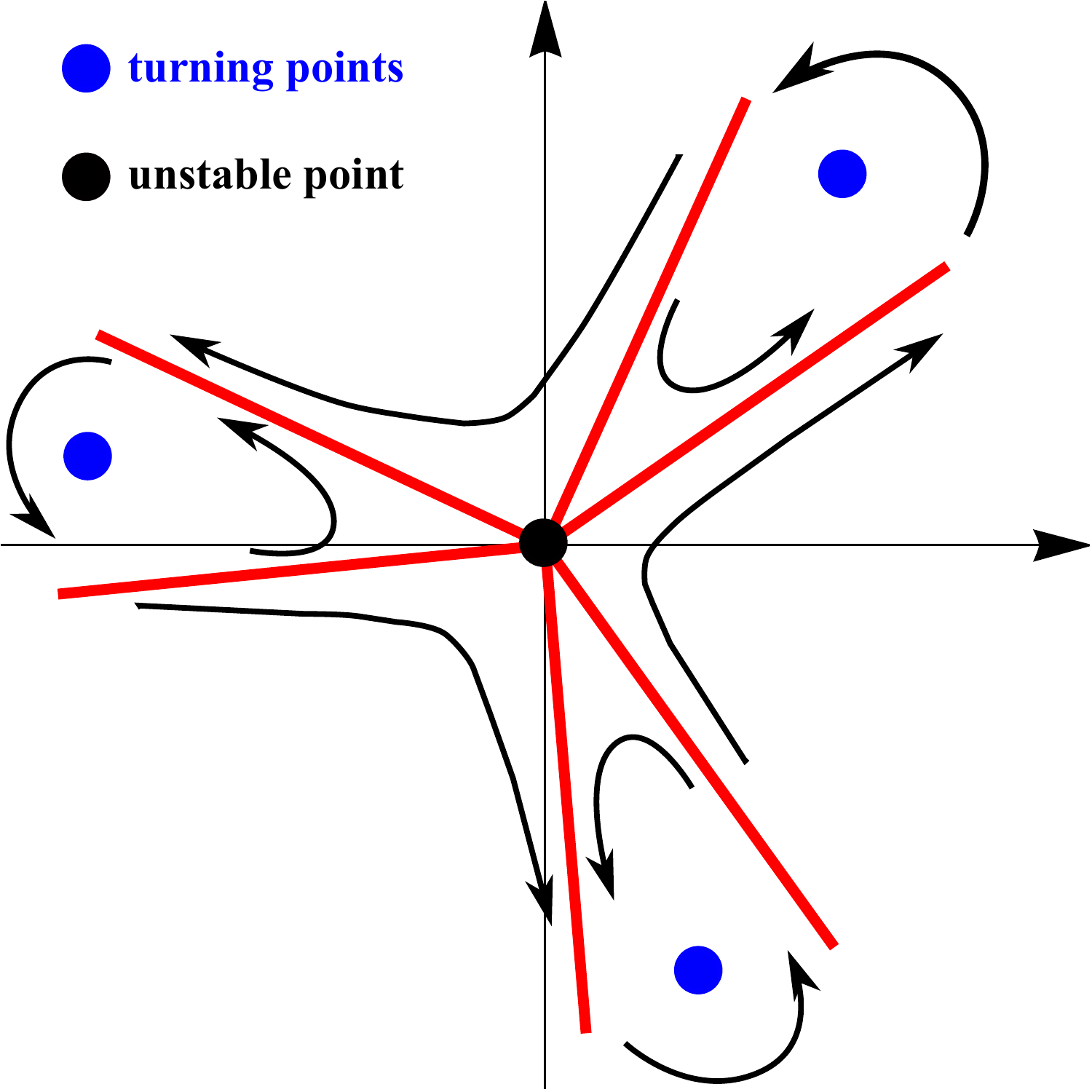}  
  \caption{sketch of the equi-energy orbits for $2\gamma/\xi\ge 1$.}
  \label{energy_contours_q4_B}
\end{subfigure}
\caption{constant energy lines obtained at $2\gamma/\xi=1.73$ by including the $q^4$ order correction in $H_{\bK}(\bq)$. The values of $\mu$ and $\nu$ are computed by evaluating the expectation values in Eq. \eqn{4rth_coefficients} numerically.}
\label{energy_contours_q4}
\end{figure}
In the regime $2\gamma/\xi\ge1$ the iso-energy contours are open and the semiclassical quantization of the orbits breaks down \cite{PhysRevB.97.144422}.
Interestingly, the $q^4$ order correction $H^{(4)}_{\bK}(\bq)$ provides new turning points away from $\bK$ around which the orbits are closed. 
This is shown in Fig. \ref{energy_contours_q4} where we represent the iso-energy contours obtained by including $q^4$ terms in the $\bk\cdot\bp$ Hamiltonian.
For this reason, in the regime $2\gamma\ge\xi$, the fourth order terms obtained from the expression
\be
H^{(4)}_{\bK}=\sum_{z,z^\prime}\ket{\Phi_z}\bra{\Phi_z}\delta H(\hat{\bm{\pi}})\,\mathcal{P}\,\delta H(\hat{\bm{\pi}})\,\mathcal{P}\,\delta H(\hat{\bm{\pi}})\,\mathcal{P}\,\delta H(\hat{\bm{\pi}})\ket{\Phi_{z^\prime}}\bra{\Phi_{z^\prime}}
\ee
are crucial for regularizing the semiclassical orbits.
By employing symmetry arguments we find that
\be
\label{4rth}
H^{(4)}_{\bK}=-i\frac{4\mu}{g^4}\left[\tau^+\otimes a\,a\,a\,a-\tau^-\otimes a^\dagger\,a^\dagger\,a^\dagger\,a^\dagger\right]-i\frac{2\nu}{g^4}\left[\tau^+\otimes\left(a^\dagger a^\dagger a a^\dagger+a^\dagger a a^\dagger a^\dagger\right)-\tau^+\otimes\left(a a a^\dagger a + a a^\dagger a a\right)\right].
\ee
For the sake of completeness, we also report the expression of the fourth order parameters $\nu$ and $\mu$: 
\bal
\label{4rth_coefficients}
&\mu=\text{Im}\left[\bra{\Phi_{\omega^*}}\,\Sigma^-\,\mathcal{P}\,\Sigma^-\,\mathcal{P}\,\Sigma^-\,\mathcal{P}\,\Sigma^-\,\ket{\Phi_\omega}\right],\\
&\nu=-\text{Im}\Bigg[\bra{\Phi_{\omega^*}}\,\Sigma^+\,\mathcal{P}\,\Sigma^+\,\mathcal{P}\,\Sigma^+\,\mathcal{P}\,\Sigma^- + \Sigma^+\,\mathcal{P}\,\Sigma^+\,\mathcal{P}\,\Sigma^-\,\mathcal{P}\,\Sigma^+ + \Sigma^+\,\mathcal{P}\,\Sigma^-\,\mathcal{P}\,\Sigma^+\,\mathcal{P}\,\Sigma^+ \\
&+ \Sigma^-\,\mathcal{P}\,\Sigma^+\,\mathcal{P}\,\Sigma^+\,\mathcal{P}\,\Sigma^+ \ket{\Phi_\omega}\Bigg].
\eal

\subsection*{Landau levels at the HOVHS}

\begin{figure}
\begin{center}
\includegraphics[width=0.6\textwidth]{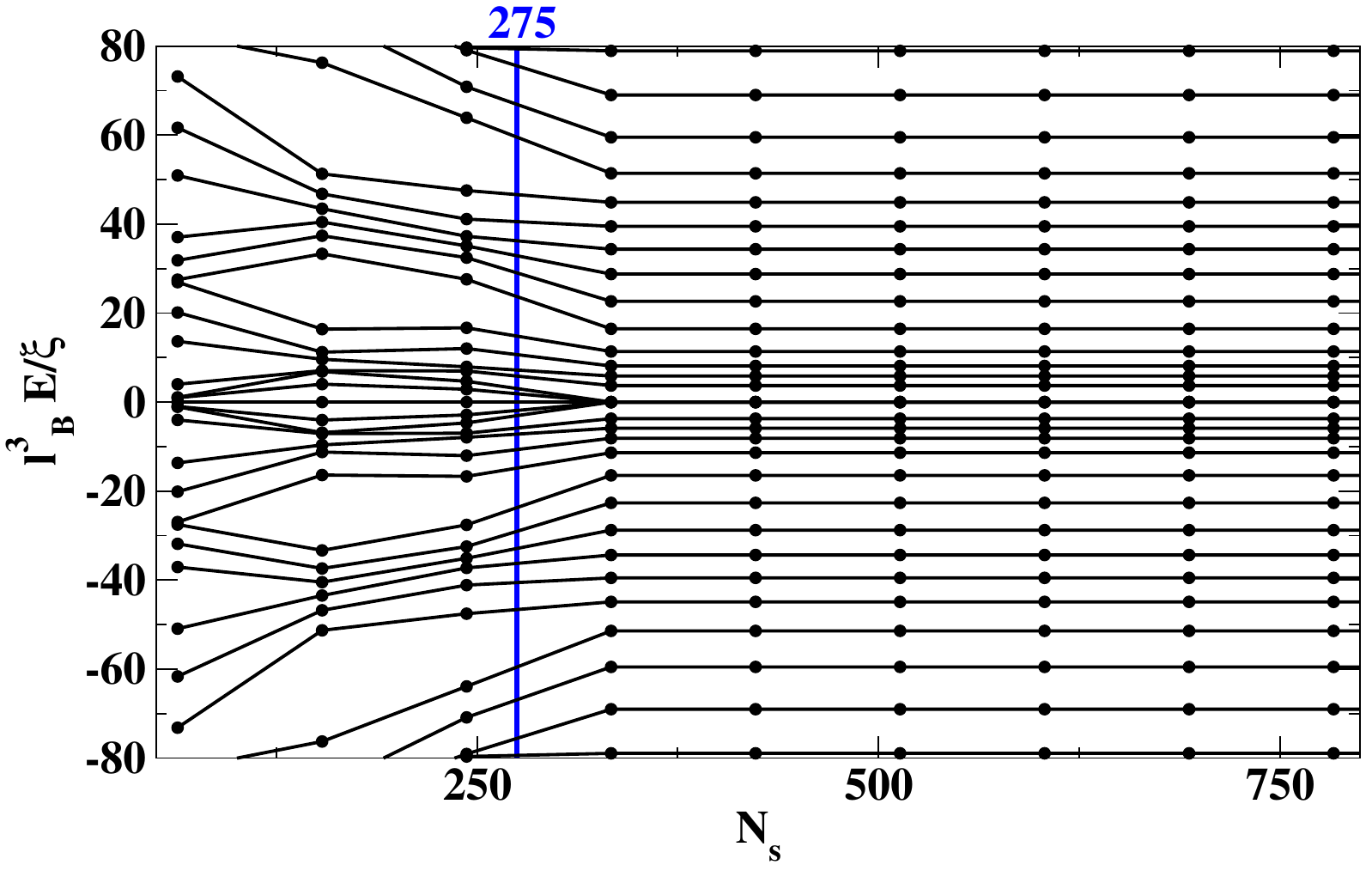}
\caption{Landau level energy as a function of the number of harmonic oscillator states $N_s$ for $2\gamma/\xi=1.73$, $\mu/\xi=0.34$ and $\nu/\xi=1.93$. Vertical blue line shows the lower bound $g^2\simeq275$.}
\label{LLS_convergence}
\end{center}
\end{figure}

\begin{figure}
\begin{center}
\includegraphics[width=0.6\textwidth]{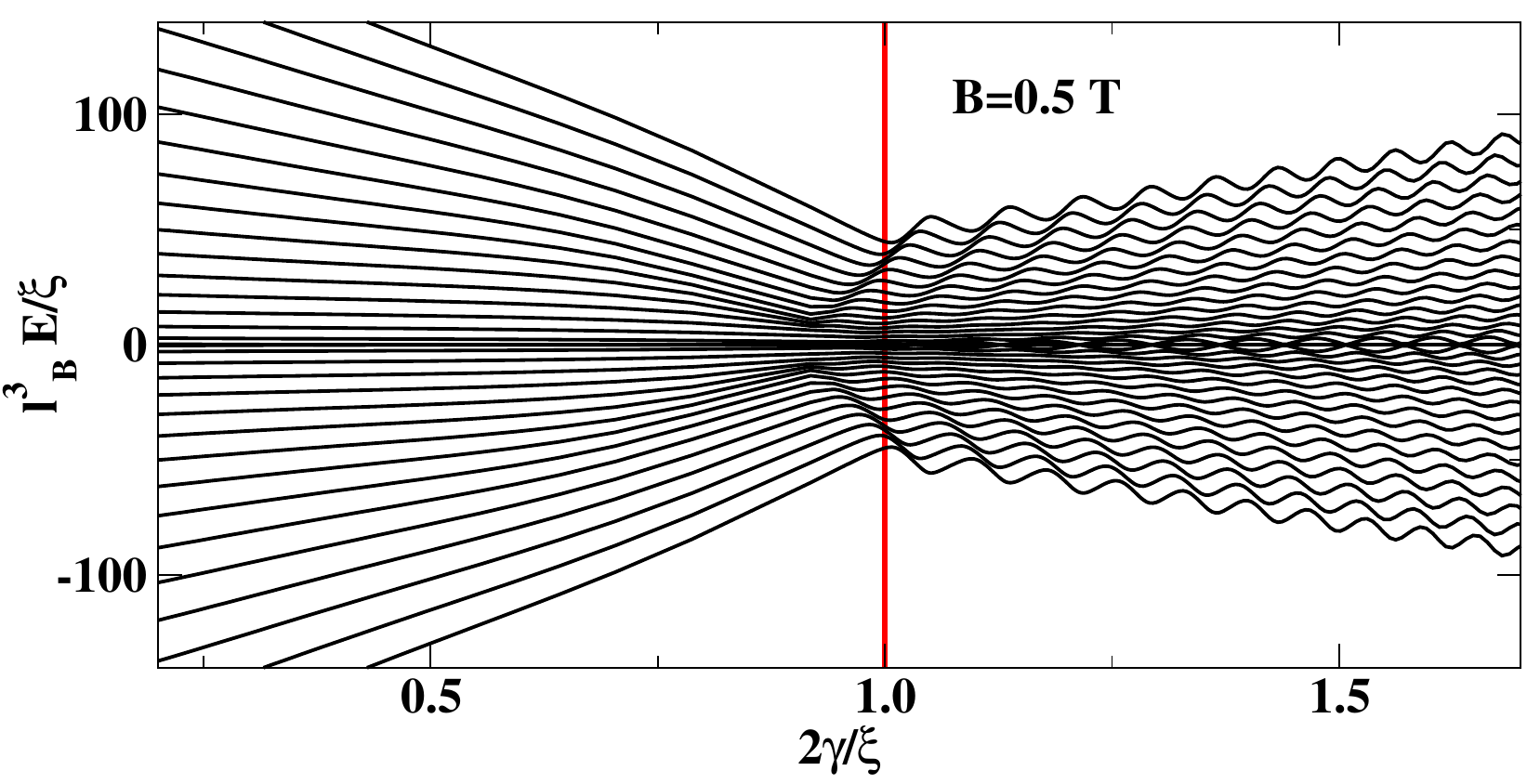}
\caption{Landau level spectrum as a function of $2\gamma/\xi$ for $B=0.5\,\text{T}$. In the plot we show only the lowest 36 energy states. The calculation has been performed by keeping $N_s=393$ harmonic oscillator states for $\ket{\Phi_\omega}$ and $\ket{\Phi_{\omega^*}}$.}
\label{LLs_spectrum}
\end{center}
\end{figure}

At the higher-order VHS ($v=\eta=0$) the $\bk\cdot\bp$ Hamiltonian~\eqn{kp_H} in a perpendicular magnetic field is
\bal
\label{3rd_4rt_H}
g^3\,H_{\bK}=&i2\sqrt{2}\gamma\sum_{n=0}^{\infty}\sqrt{(n+1)(n+2)(n+3)} \left[\tau^0\otimes\ket{n}\bra{n+3}-\tau^0\otimes\ket{n+3}\bra{n}\right]\\
&+i2\sqrt{2}\xi\sum_{n=0}^{\infty}(n+1)^{3/2}\left[\tau^- \otimes\ket{n+1}\bra{n}-\tau^+\otimes\ket{n}\bra{n+1}\right]\\
&-i\frac{2\nu}{g}\sum_{n=0}^{\infty}\sqrt{(n+1)(n+2)}(2n+3)\left[\tau^+\otimes\ket{n+2}\bra{n}-\tau^-\otimes\ket{n}\bra{n+2}\right]\\
&-i\frac{4\mu}{g}\sum_{n=0}^{\infty}\sqrt{(n+1)(n+2)(n+3)(n+4)}\left[\tau^+\otimes\ket{n}\bra{n+4}-\tau^-\otimes\ket{n+4}\bra{n}\right].
\eal
where we have projected the terms \eqn{3rd} and \eqn{4rth} in the basis of eigenstates of the harmonic oscillator $\ket{n}$ with $n=0,1,\cdots$. 
The fourth order contribution is reduced by a factor $1/g$ with respect to the third order one. 
For a magnetic field that ranges from $B=0.005-1.0\,\text{T}$ and in the magic-angle region $k_\theta\simeq0.47\,\text{nm}^{-1}$ ($L_s\simeq8.8\,\text{nm}$) the value of $g$ goes from $\sim170$ to $\sim12$. In order to find the spectrum of the Hamiltonian in Eq. \eqn{3rd_4rt_H} we truncate the basis of harmonic oscillator states to a given integer $N_{s}$, i.e. $n=0,1,\cdots,N_s-1$. The number of states $N_s$ needed to sample the turning points produced by the $q^4$ correction goes as $N_s\sim(k_\theta l_B)^2= g^2$. 
Therefore, we conclude that $N_s$ increases like $1/B$ as we decreases the value of $B$. 
For a magnetic field of $B=0.5\,\text{T}$ the value of $g$ in the magic angle region is $g\simeq 16.6$ and we expect to achieve convergence for $N_s> 275$. 
This is confirmed by Fig. \ref{LLS_convergence} where we show the Landau level energy as a function of $N_s$ for $2\gamma/\xi=1.73$. 
Finally, Fig. \ref{LLs_spectrum} shows the Landau level spectrum across the Lifshitz transition $2\gamma/\xi=1$ for a magnetic field $B=0.5\,\text{T}$.

\bibliographystyle{apsrev}
\bibliography{SMbiblio}